\documentclass[fleqn,usenatbib]{mnras}

\usepackage{newtxtext,newtxmath}
\usepackage[T1]{fontenc}
\usepackage{ae,aecompl}
\usepackage{pdflscape}

\usepackage{graphicx}	% Including figure files
\usepackage{amsmath}	% Advanced maths commands
\usepackage{amssymb}	% Extra maths symbols
\usepackage{longtable}
\usepackage{multirow}
\title[The 3D shape of galactic bars]{The intrinsic three-dimensional shape of galactic bars}

\author[J. M\'endez-Abreu]{J. M\'endez-Abreu$^{1,2}$\thanks{E-mail: jairomendezabreu@gmail.com},
L. Costantin$^{3}$,
J. A. L. Aguerri$^{1,2}$,
A. de Lorenzo-C\'aceres$^{1,2}$,
\newauthor E. M. Corsini$^{3,4}$
\\
$^{1}$Instituto de Astrof\'isica de Canarias, Calle V\'ia L\'actea s/n, E-38205 La Laguna, Tenerife, Spain\\
$^{2}$Departamento de Astrof\'isica, Universidad de La Laguna, E-38200 La Laguna, Tenerife, Spain\\
$^{3}$Dipartimento di Fisica e Astronomia 'G. Galilei', Universit\'a di Padova, vicolo dell'Osservatorio 3, I-35122 Padova, Italy\\
$^{4}$INAF-Osservatorio Astronomico di Padova, vicolo dell'Osservatorio 5, I-35122 Padova, Italy\\
}

\date{Accepted XXX. Received YYY; in original form ZZZ}

\pubyear{2018}

\begin{document}
\label{firstpage}
\pagerange{\pageref{firstpage}--\pageref{lastpage}}
\maketitle

% Abstract of the paper
\begin{abstract}
%Stellar bars  are thought to play  an essential role in  their secular
%evolution. However,  their three dimensional (3D)  intrinsic shape has
%been barely studied observationally.
%
We   present   the   first   statistical  study   on   the   intrinsic
three-dimensional (3D) shape of a sample of 83 galactic bars extracted
from the CALIFA survey.
We  use the  galaXYZ code  to derive  the bar  intrinsic shape  with a
statistical approach.  The method  uses only the geometric information
(ellipticities and position angles) of  bars and discs obtained from a
multi-component    photometric    decomposition    of    the    galaxy
surface-brightness distributions.
We  find  that  bars  are  predominantly  prolate-triaxial  ellipsoids
(68\%),   with  a   small  fraction   of  oblate-triaxial   ellipsoids
(32\%). The typical flattening (intrinsic $C/A$ semiaxis ratio) of the
bars in our  sample is 0.34, which matches well  the typical intrinsic
flattening of  stellar discs at  these galaxy masses.   We demonstrate
that, for prolate-triaxial  bars, the intrinsic shape  of bars depends
on  the galaxy  Hubble  type  and stellar  mass  (bars  in massive  S0
galaxies  are thicker  and more  circular than  those in  less massive
spirals).  The  bar intrinsic shape  correlates with bulge,  disc, and
bar  parameters.   In  particular   with  the  bulge-to-total  ($B/T$)
luminosity ratio, disc $g-r$ color,  and central surface brightness of
the  bar,  confirming the  tight  link  between  bars and  their  host
galaxies.  Combining  the probability  distributions of  the intrinsic
shape of  bulges and bars  in our sample we  show that 52\%  (16\%) of
bulges are  thicker (flatter)  than the  surrounding bar  at 1$\sigma$
level.
We  suggest that  these  percentages might  be  representative of  the
fraction   of  classical   and   disc-like  bulges   in  our   sample,
respectively.
\end{abstract}

% Select between one and six entries from the list of approved keywords.
% Don't make up new ones.

%%%%%%%%%%%%%%%%%%%%%%%%%%%%%%%%%%%%%%%%%%%%%%%%%%

%%%%%%%%%%%%%%%%% BODY OF PAPER %%%%%%%%%%%%%%%%%%
%--------------------------------------------------------
%--------------------------------------------------------
\section{Introduction}
\label{sec:intro}

Stellar  bars are   common  structures  in disc  galaxies in  the
nearby Universe \citep{marinovajogee07,  barazza08,aguerri09} and they
are considered the  main internal mechanism driving  the dynamical and
secular evolution  of disc galaxies  \citep{kormendykennicutt04}.  The
presence of  a bar  is able  to modify the  external appeareance  of a
galaxy,  changing  its structure  and  morphology  within the  central
$\sim$ 10  kpc.  Numerical  simulations have  found that  stellar bars
redistribute the angular momentum between the baryonic and dark matter
components   \citep{debattistasellwood98,debattistasellwood00}.    The
relative amount of  exchanged angular momentum is  related to specific
properties of  the galaxies, such as  the bar mass, halo  density, and
halo                        velocity                        dispersion
\citep{athanassoula03,sellwooddebattista06,athanassoula13}.         In
addition, bars  are able to  funnel material toward the  galaxy center
where starbursts can ignite  \citep{sheth05,shlosman90}, and they have
been proposed  as an efficient  mechanism to create new  structures in
the      galaxy     centers      such     as      disc-like     bulges
\citep{kormendykennicutt04},       inner      star-forming       rings
\citep{buta03,munoztunon04},          and          inner          bars
\citep{erwin04,debattistashen07,delorenzocaceres13}.        Therefore,
understanding the mechanisms  leading to the formation of  bars is key
to understand galaxy evolution in general.

The observed properties  of stellar bars, and their  relation with the
characteristics of the host galaxies have been extensively reviewed in
the literature.   It is now established  that the main driver  for the
presence   of  a   bar  is   the  galaxy   mass  \citep{mendezabreu10,
  nairabraham10,erwin17b}, with  the bar frequency reaching  a peak at
$M_{\star}  \sim  10^{9.5}M_{\sun}$   \citep{mendezabreu12}  and  quickly
declining towards both lower and higher masses. Many observational and
theoretical studies have also investigated other internal and external
galaxy properties to understand which galaxies are more or less likely
to     host    bars     obtaining    some     contradictory    results
\citep{barazza09,athanassoula13,diazgarcia16,martinezvalpuesta17}.  In
\citet{aguerri09}, we found that  red, massive, gas-poor galaxies host
less   and   shorter   bars   than  blue,   low-mass,   and   gas-rich
galaxies.   These  results   were  similar   to  those   presented  in
\citet{barazza08} and  \citet{nairabraham10}.  However,  other authors
obtained     completely     opposite     results     \citep{masters11,
  skibba12,consolandi16}.  In \citet{mendezabreu12}  we discussed that
most of  these discrepancies might  be solved if galaxy  samples would
have been carefully  selected in both stellar mass  and environment to
avoid   biases   when   dealing  with   bar   statistics.    Recently,
\citet{erwin17b} confirmed this argumentation  and he also pointed out
the problems  of dealing with  poor spatial resolution  when analysing
galaxy bars.

Despite the  tremendous progress in  the field, an important  piece of
information is  still hidden  in the intrinsic  three-dimensional (3D)
structure of  stellar bars, and  how the bar  shape is related  to the
different galaxy  properties.  $N$-body simulations  have demonstrated
how the 3D structure of stellar bars is strongly dependent on the time
since  bar formation.   Bars  form  spontaneously in  self-gravitating
rotating                         galactic                        discs
\citep{combes90,debattistasellwood00,athanassoula03};  initially  they
have a  thin vertical  density profile  similar to  that of  the disc,
i.e.,  bars  are  formed  by   the  re-arrangement  of  disc  material
\citep{athanassoula13}.  This scenario have  been confirmed by orbital
analysis where  planar and circular  orbits become more  elongated and
material  gets trapped  around the  stable periodic  orbits of  the x1
family \citep{contopoulos80,athanassoula83, athanassoula92}.  However,
this configuration quickly changes due to  the bar buckling out of the
disc plane,  which modifies  the vertical structure  of the  bar inner
regions  and it  creates a  substantially thicker  component than  the
surrounding  disc  \citep{combessanders81,raha91,martinezvalpuesta06}.
This vertical structure, which appears  as boxy- or peanut-shaped when
the galaxy is  seen edge-on, has also been demonstrated  to be part of
the  bar through  3D orbital  analysis \citep{skokos02}.   In summary,
numerical simulations suggest that, on the long term, most bars should
be formed by  an inner thick component and an  outer part thinner than
the inner region \citep{athanassoula05}.

Observationally, boxy/peanut  (B/P) shaped bars have  been detected in
many  studies of  edge-on and,  indirectly, also  in face-on  galaxies
\citep{lutticke00,yoshino15,mendezabreu08b,erwindebattista13,
  ciamburgraham16}. However, they have  not been generally included as
a separate  component in either photometric  decompositions \citep[but
  see][and  references  therein]{laurikainen16}  or  bar  deprojection
schemes  \citep{martin95,gadotti07}.   In  fact,  \citet{zou14}  using
numerical  simulations  to  test  their  2D  bar  deprojection  method
concluded  that, to  first order  approximation, one-ellipsoid  models
could represent the 3D structure of  the bar reasonably well. This was
also pointed out in early hydrodnamical simulations of barred galaxies
\citep{hunter88,england89,aguerri01}.  In  addition, the  peanut shape
generally does not  encompass the full extent of the  bar, i.e., there
is    some    flat    bar    outside   the    buckled    inner    part
\citep{lutticke00,mendezabreu08b}.    \citet{erwindebattista13}   also
argue that  not every  bar thickens vertically  and estimated  that at
least 13\% of bars in galaxies have not buckled.

In this paper, we present the first statistical study of the intrinsic
shape of  stellar bars  using the  observed sample  of the  Calar Alto
Legacy    Integral    Field    Area     survey    Data    Release    3
\citep[CALIFA-DR3][]{sanchez16}.  We use the galaXYZ code described in
\citet{mendezabreu10}  and \citet{costantin18}  and  compare the  bar
intrinsic shape with other observed galaxy properties to shed light on
the formation  of these structures. To  this aim, we consider  bars to
be,  at   first  order,  single  triaxial   ellipsoids.   Despite  the
limitations of such a description, it is an appropriate starting point
to study in  a quantitative way the  3D shape of bars  with respect to
galaxy   properties.     The   paper   is   organised    as   follows:
Sect.~\ref{sec:sample} describes  the sample of galaxies  used in this
work.  Sect.~\ref{sec:bars3D}  details the  methodology to  derive the
intrinsic  shape  of stellar  bars  from  their projected  photometric
properties.   Sect.~\ref{sec:results} describes  the  main results  of
this paper  about the intrinsic  shape of  bars and its  relation with
other galaxy properties.  The discussion of the results in the context
of  bar  formation  is  also  done  in  Sect.~\ref{sec:results}.   The
conclusions are given  in Sect.~\ref{sec:conclusions}.  Throughout the
paper  we  assume  a  flat  cosmology with  $\Omega_{\rm  m}$  =  0.3,
$\Omega_{\rm \Lambda}$  = 0.7,  and a  Hubble constant  $H_0$ =  70 km
s$^{-1}$ Mpc$^{-1}$.

%--------------------------------------------------------
%--------------------------------------------------------
\section{CALIFA sample of barred galaxies}
\label{sec:sample}

The    sample     of    barred    galaxies    was     selected    from
\citet{mendezabreu17}. They present  a two-dimensional multi-component
photometric  decomposition   of  404  galaxies  from   the  CALIFA-DR3
\citep[][]{sanchez16}.   They represent  all galaxies  with either  no
clear signs  of interaction  or not strongly  inclined within  the 667
galaxies observed  in the final  CALIFA data release.  They  found 162
barred galaxies  out of 404 galaxies.   Following \citet{costantin18},
we impose  an inclination  constraint to  galaxies with  $25^{\circ} <
\theta <  65^{\circ}$.  This assures  the robustness of our  method to
derive the  intrinsic shape of our  bar sample.  We ended  up with 125
barred  galaxies  meeting  this  criteria.  We  further  discarded  42
galaxies  for   the  following  reasons:  i)   galaxies  with  complex
morphology in the  outer parts of the discs, e.g.,  strong spiral arms
or  asymmetric  discs,  that  make  it  difficult  to  deriving  their
geometric (ellipticity  and position angle) properties  (14 galaxies),
ii)  galaxies with  very low  surface brightness  discs ($\mu_0  > 21$
mag/arcsec$^2$)  or bars  ($\mu_{0,\rm  bar}  > 22.5$  mag/arcsec$^2$)
making it  difficult to  measuring their  geometry (8  galaxies), iii)
galaxy discs for which the  ellipticity and/or position angle from the
photometric  decomposition does  not  match the  outer  values of  the
corresponding radial  profiles derived from the  isophotal analysis (9
galaxies), iv) galaxies  with an offcentered bar (3  galaxies), and v)
galaxies for  which the uncertainties  in the derived  intrinsic shape
($B/A$ or  $C/A$) are larger  than 1  (8 galaxies).  The  final sample
amounts to 83 galaxies.

\begin{figure*}
\includegraphics[angle=90,width=0.9\textwidth]{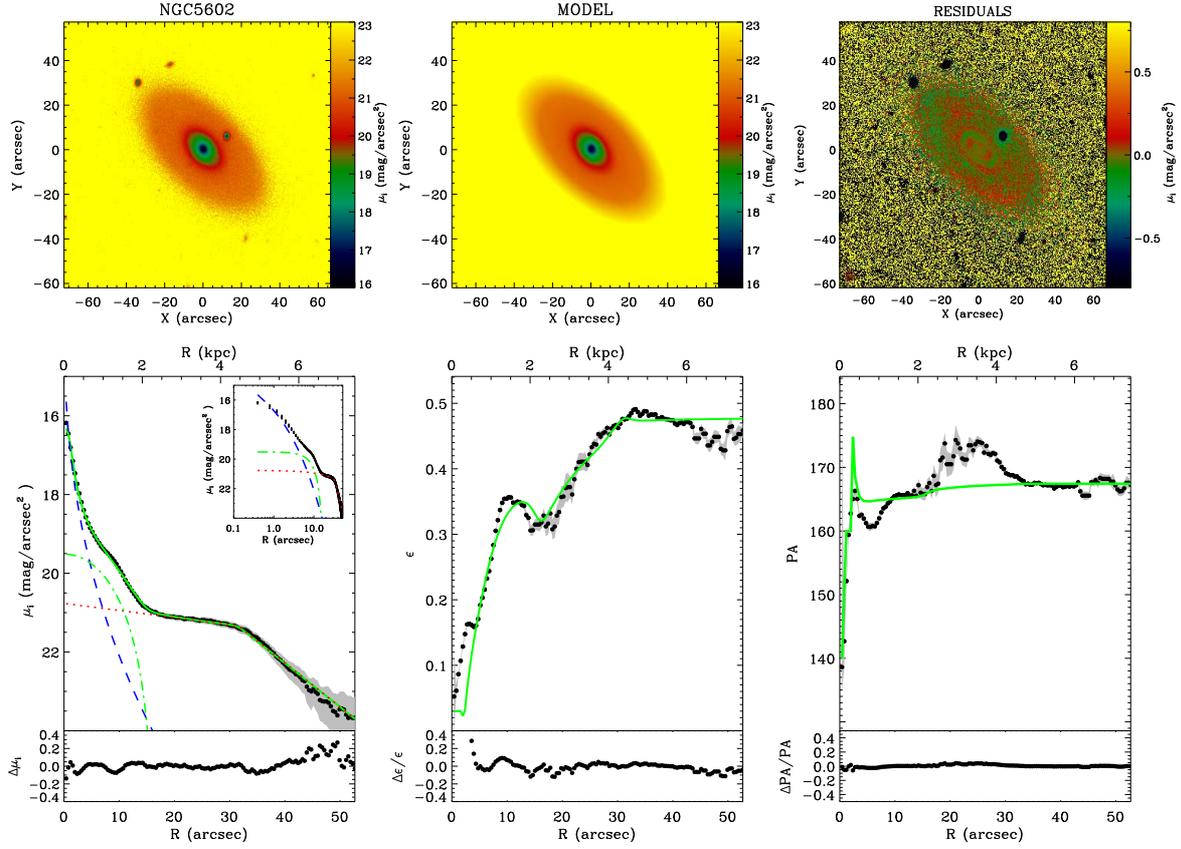}
%\vspace{0.3cm}
\caption{Example  of the  2D  photometric  decomposition analysis  for
  NGC~5602.   The  figure shows  the  best  fit obtained  using  three
  components (bulge, bar,  and truncated disc) for  the $i-$band.  Top
  left panel: galaxy  image.  Top middle panel:  best-fitting model of
  the  galaxy image.   Top  right panel:  residual  image obtained  by
  subtracting the best-fit  model from the galaxy  image.  Bottom left
  panel:  ellipse-averaged surface  brightness radial  profile of  the
  galaxy (black dots) and best-fit model (cyan solid line).  The light
  contributions  of the  bulge (dashed  red line),  disc (dotted  blue
  line),  and bar  (dotted-dashed green  line) are  shown.  The  upper
  inset shows  a zoom of  the surface-brightness  data and fit  with a
  logarithmic scale for the distance to  the center of the galaxy.  1D
  surface brightness residuals (in  mag/arcsec$^2$ units) are shown in
  the bottom sub-panel.  Bottom  middle panel: ellipse-averaged radial
  profile of ellipticity of the galaxy (black dots) and best-fit model
  (cyan solid  line).  1D residuals  (in percentage) are shown  in the
  bottom  sub-panel.   Bottom  right  panel:  ellipse-averaged  radial
  profile of  position angle of  the galaxy (black dots)  and best-fit
  model (cyan solid line).  1D  residuals (in percentage) are shown in
  the bottom  sub-panel. The  grey shaded areas  in the  bottom panels
  represent the measurement errors derived from the {\tt ellipse IRAF}
  task when applied to the galaxy image.}
\label{fig:phot}
\end{figure*}

\begin{figure*}
\includegraphics[bb=154 54 378 738,angle=90,width=0.9\textwidth]{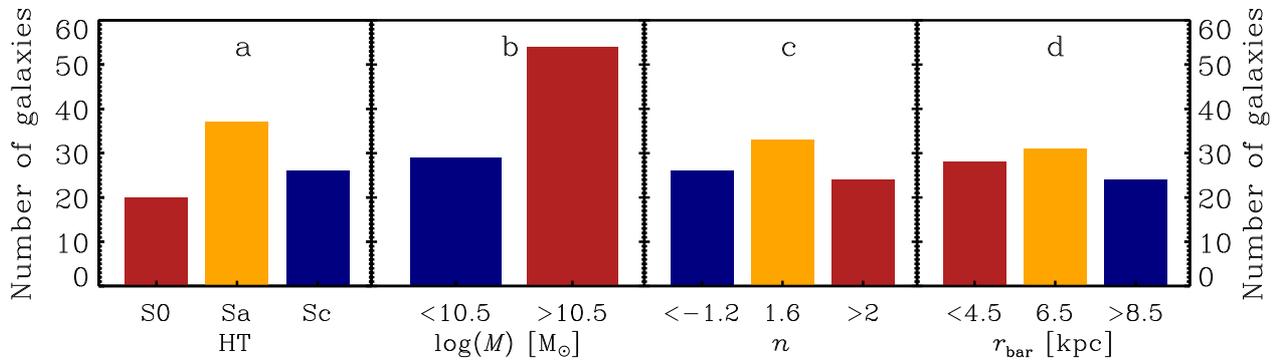}
%\vspace{0.3cm}
\caption{Distribution  of the  galaxy  Hubble type  (panel  a; Sa  bin
  comprises Sa-Sab-Sb galaxies, while Sc bin comprises Sbc-Scd-Sd-Sdm
  galaxies),  stellar mass  (panel  b), S\'ersic  index  of the  bulge
  (panel c), and bar radius (panel d). The galaxy properties are taken
  from \citet{mendezabreu17} and \citet{walcher14}.}
\label{fig:properties}
\end{figure*}

\citet{mendezabreu17}  carried   out  the   photometric  decomposition
analysis using the  $g-$, $r-$, and $i-$band images.  In  this work we
have used the  results from the $i-$band images to  better resolve the
bar component  minimising the dust  effects with respect to  the other
SDSS passbands.  The average $i-$band point spread function (PSF) full
width at half maximum (FWHM) of  the images is 1.1$\pm$0.2 arcsec with
a typical  depth of $\mu_i \sim$  26 mag arcsec$^{-2}$. An  example of
the   2D   photometric  decomposition   for   NGC5602   is  shown   in
Fig.~\ref{fig:phot}.  The galaxies main  properties (i.e., Hubble type
$HT$, stellar mass $M_{\star}$, S\'ersic  index parameter $n$, and bar
radius $r_{\rm bar}$) from \citet{mendezabreu17} and \citet{walcher14}
are shown in Fig.~\ref{fig:properties}.

%--------------------------------------------------------
%--------------------------------------------------------

%--------------------------------------------------------
%--------------------------------------------------------

%--------------------------------------------------------
%--------------------------------------------------------
\section{Bars intrinsic shape}
\label{sec:bars3D}

The derivation  of the intrinsic shape  of the individual bars  in our
sample  was  performed  using  the   galaXYZ  code.  This  method  has
previously been applied  to the analysis of the intrinsic  3D shape of
galactic  bulges \citep{mendezabreu10,  costantin18}. However,  it is
worth  noticing that  our statistical  approach is  applicable to  any
galactic structure if the initial assumptions are fulfilled.

For the sake  of clarity, we describe here the  main hypotheses of our
method. In  order to  characterise the  intrinsic shape  of a  bar, we
first  assumed that  it can  be  successfully modelled  by a  triaxial
ellipsoid that shares the same equatorial plane as the disc. Secondly,
the  galaxy disc  is  considered to  be an  oblate  ellipsoid with  an
intrinsic thickness given by a  normal distribution function with mean
intrinsic axial ratio  $\langle q_{0,d} \rangle$ =  0.267 and standard
deviation      $\sigma_{q_{0,       {\rm      d}}}$       =      0.102
\citep{rodriguezpadilla13}.  Moreover, a  third condition imposes that
the  bar and  disc share  the same  center, which  coincides with  the
galaxy  center. This  condition  is also  imposed  in the  photometric
decomposition where all components are   forced to share the same
center.  We also visually checked  the individual galaxies to look for
offcentered  bars such  as those  presented in  \citet{kruk17} finding
only  3 out  of  125. This  low  fraction of  offcentered  bars is  in
agreement  with the  findings  of \citet{kruk17}  that  most of  these
systems are located in low-mass  galaxies.  They were removed from the
sample (see Sect.~\ref{sec:sample}).

   Regarding   the    first   hypothesis,    we   discussed    in
  Sect.~\ref{sec:intro} about  the possibility of bars  being composed
  by both an  inner vertically thick component  (usually associated to
  the  presence of  a B/P  structure)  and another  more extended  and
  thinner  bar.   In  our  photometric decomposition,  the  inner  B/P
  structure is not  included as a different  analytical component, and
  we  visually checked  all the  fits to  assure the  external bar  is
  fitted.   We  thus  consider  our  intrinsic  shape  results  to  be
  representative of the outer extended  bar.  However, the presence of
  a B/P  structure might  still be affecting  the measurements  of the
  geometric parameters involved in our analysis of the 3D shape, i.e.,
  the  bar and  disc  ellipticities and  position  angles.  To  better
  quantify  the effect  of possible  B/P structures  in our  sample we
  carried out a visual classification  of our sample galaxies in order
  to  detect  the  possible  presence of  inner  B/P  structures.   We
  followed  the  criteria  established  in   a  series  of  papers  by
  \citet{laurikainen11,athanassoula15,laurikainensalo17}:  a potential
  B/P  is detected  based  on  the presence  of  a  close to  circular
  isophotal  contour in  the  central  part of  the  bar with  smaller
  ellipticity compared  to the outer  bar.  In addition, we  also used
  the criteria proposed by \citet{erwindebattista13,erwindebattista17}
  to detect  B/P structures.  This  is based  on the presence  of boxy
  isophotes in the  inner part of the bar, accompanied  by narrow, and
  offset with respect  to the major axis, isophotes in  the outer bar.
  These features are called spurs.  A good example of the two types of
  projections  expected  for B/P  structures  is  given in  Fig.~1  of
  \citet{li17}.  Using this classification scheme we found that 22 and
  8  galaxies  were  classified  as barlenses  or  possible  barlenses
  (barlens?), respectively.  Similarly, we found that 3 and 3 galaxies
  were    classified    as    B/P     or    possible    B/P    (B/P?),
  respectively. Finally,  we classified 47 galaxies  as not presenting
  any characteristic of an inner B/P structure. The classification for
  each  galaxy is  shown in  Table~\ref{tab:results}.  Separating  our
  sample into  low- ($\log(M_{\star}/M_{\sun}$) < 10.5)  and high-mass
  ($\log(M_{\star}/M_{\sun}$) > 10.5) galaxies  we found 31\% and 52\%
  of our bars are buckled,  respectively.  This result is in agreement
  with previous result from \citet{li17} and \citet{erwindebattista17}
  showing an  increase on the  fraction of B/P structures  with galaxy
  mass.

 We also  studied the typical effect that a  missing B/P structure
  would have in  our results. We extensively  describe the methodology
  and results  of this  test in Appendix~\ref{sec:barlens}.   We found
  that the values of the bar intrinsic semiaxis ratios $B/A$ and $C/A$
  would be  both systematically overestimated by  0.04 when dismissing
  the  B/P.   Similarly,  the  uncertainties in  the  intrinsic  shape
  derived  by  our  method   (see  Table~\ref{tab:results})  would  be
  underestimated by  0.05 and 0.03  in $B/A$ and  $C/A$, respectively.
  These  small  biases  could  be affecting  our  galaxies  previously
  classified as hosting a boxy/peanut structure.

The methodology and equations  used to derive the intrinsic shape
  of our  bars are extensively discussed  in \citet{mendezabreu10} and
  \citet{costantin18}. We refer the readers to those papers for a full
  description of the  problem. Here, we just remind that  the scope of
  our method is to derive, starting  from a set of projected geometric
  parameters  for  the  bar   (ellipticity  $\epsilon_{\rm  bar}$  and
  position   angle  $PA_{\rm   bar}$)   and   the  disc   (ellipticity
  $\epsilon_{\rm d}$  and position angle $PA_{\rm  d}$), the intrinsic
  semiaxes ($A$, $B$, $C$) of the three-dimensional triaxial ellipsoid
  describing our bars.  The solution  to this inversion problem is not
  unique  since  we  are  missing  one  observable,  i.e.,  the  angle
  subtended by  the major axis  of the bar and  the major axis  of the
  disc  in the  galaxy  plane (Euler  $\phi$  angle).  Therefore,  our
  methodology propose a statistical approach to this problem.

\begin{figure*}
\vspace{0.3cm}
\includegraphics[bb=50 350 558 708,width=0.41\textwidth]{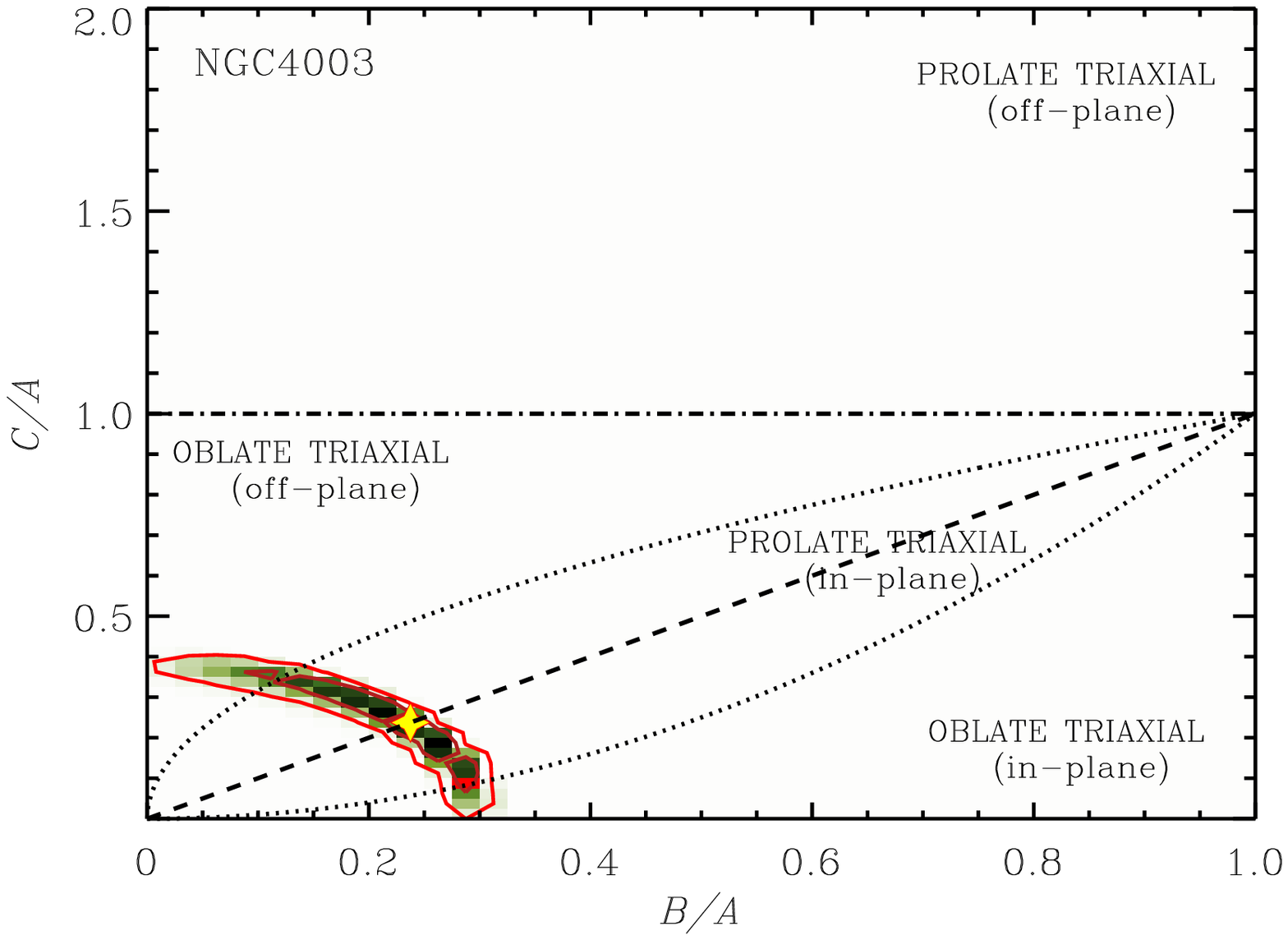}
\includegraphics[bb=50 350 558 708,width=0.41\textwidth]{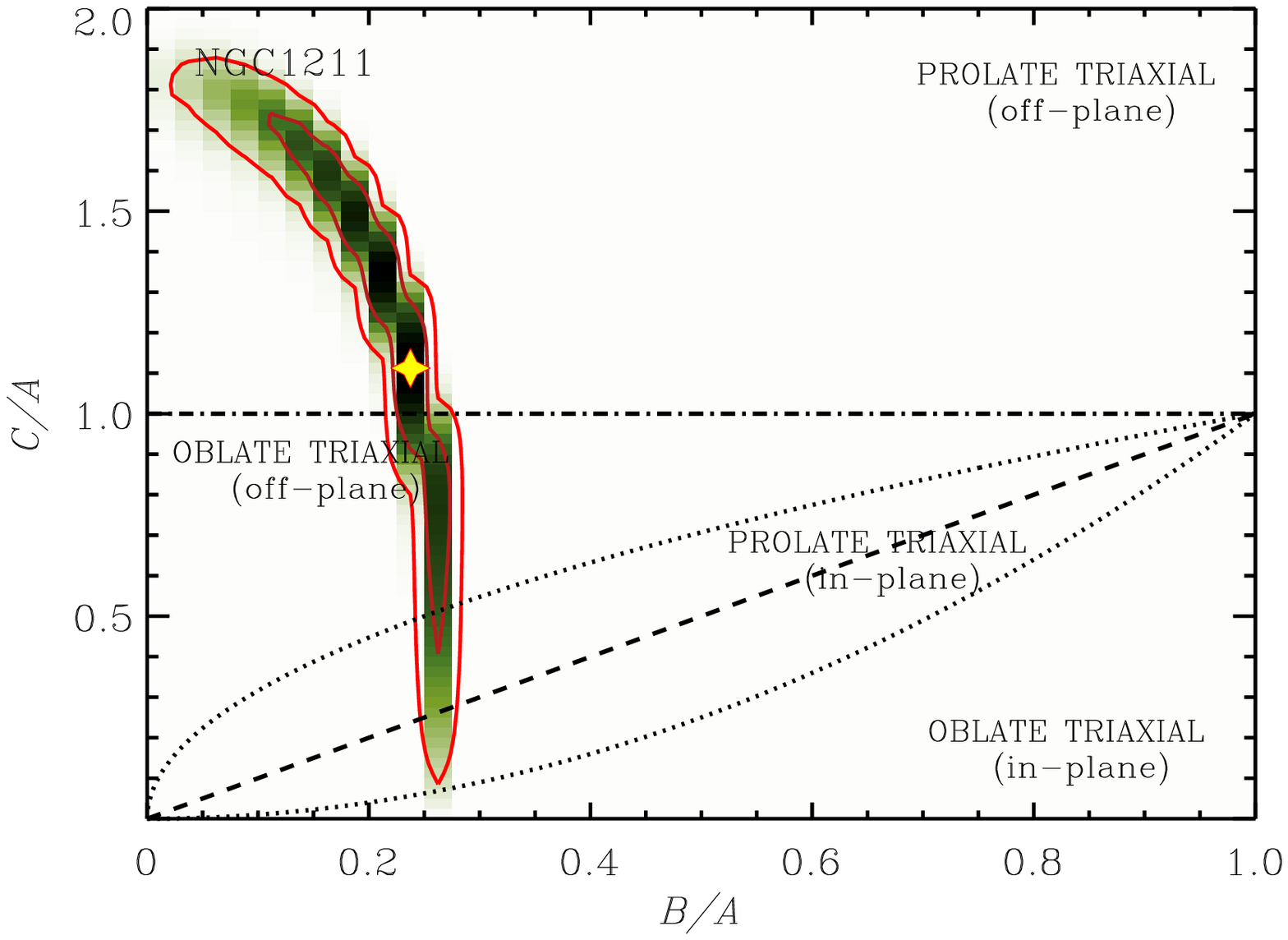}
\includegraphics[bb=50 350 558 708,width=0.41\textwidth]{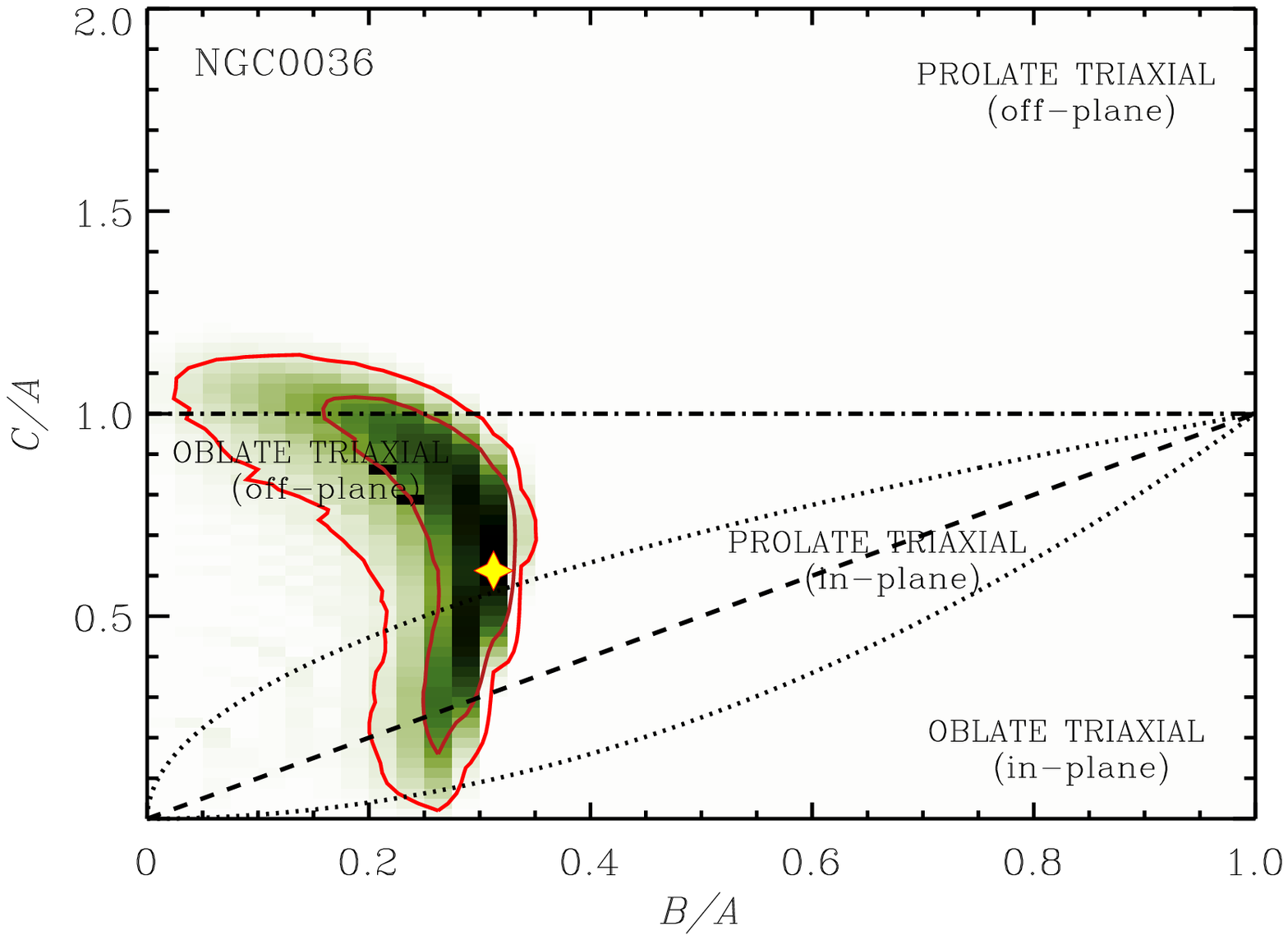}
\includegraphics[bb=50 350 558 708,width=0.41\textwidth]{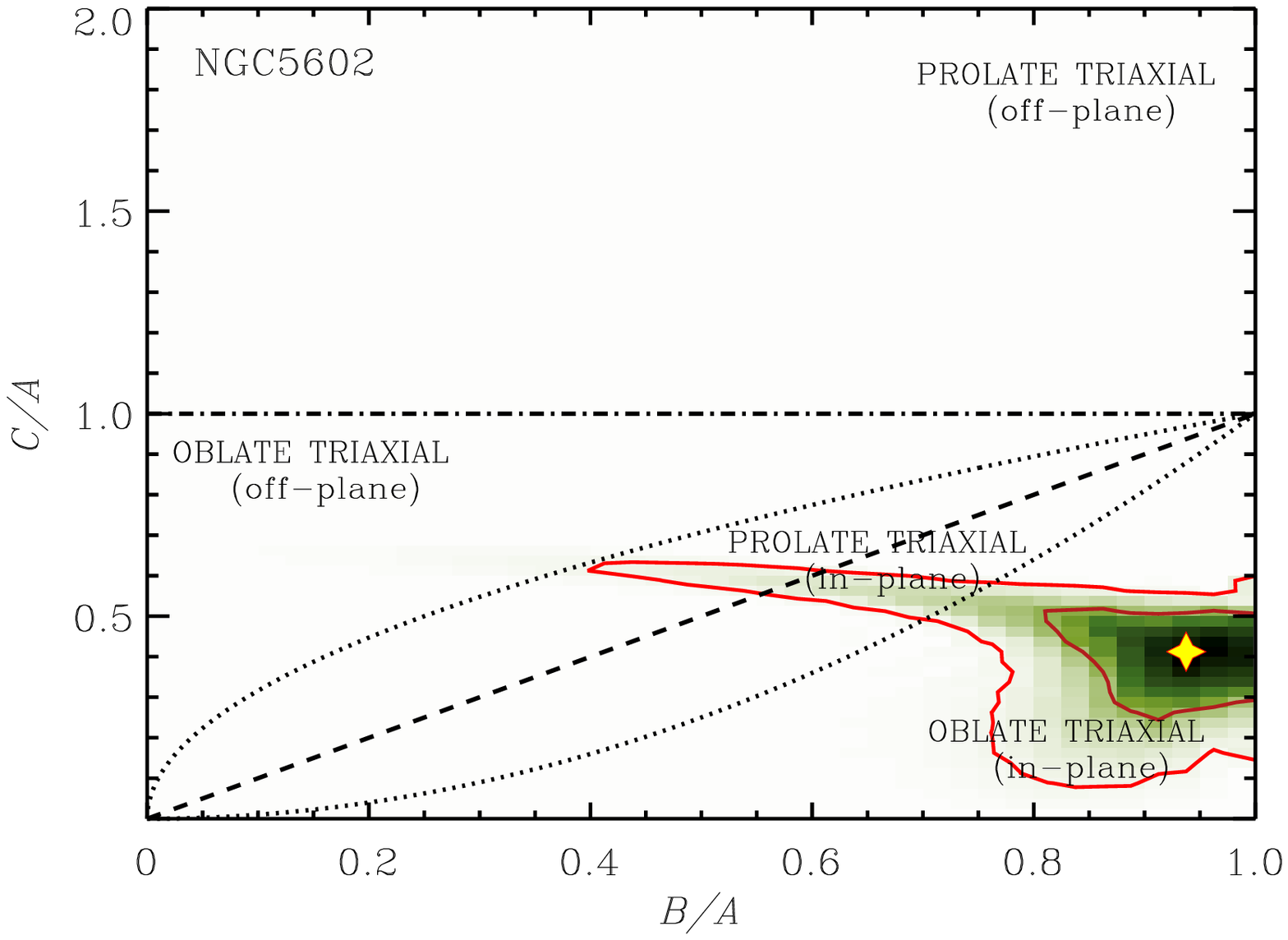}
\caption{Distribution of the intrinsic axial ratios $B/A$ and $C/A$ of
  four bars in our initial  galaxy sample. The yellow star corresponds
  to the most probable value of  $B/A$ and $C/A$.  The inner and outer
  red solid contours represent the 1$\sigma$ and 2$\sigma$ probability
  contours of the intrinsic semiaxes ratios $B/A$ and $C/A$ consistent
  with  the  bar  and  disc geometric  parameters  measured  from  our
  photometric  decomposition.  Different  lines  mark  the regimes  of
  prolate-triaxial    (off-plane),     prolate-triaxial    (in-plane),
  oblate-triaxial (off-plane), oblate-triaxial  (in-plane). We show as
  an example the four main possible  types of bars found in our sample
  covering the four  different shapes.  NGC~1211 was  removed from the
  sample due to the large uncertainties  in $C/A$ and it is shown here
  as an example of prolate-triaxial (off-plane). }
\label{fig:example3D}
\end{figure*}

Figure~\ref{fig:example3D}  shows   an  example  of   the  probability
distribution function  (PDF) for  the intrinsic semiaxis  ratios $B/A$
and $C/A$ of  some example galaxies in our sample.  It is worth noting
that the width of the distribution  in either $B/A$ and $C/A$ does not
only depend on the photometric decomposition errors, but mostly on the
lack  of   knowledge  of   the  Euler  $\phi$   angle.   Nevertheless,
Fig.~\ref{fig:example3D} shows how our statistical analysis is able to
produce important constraints on the shape of our sample bars.

%--------------------------------------------------------
%--------------------------------------------------------

%--------------------------------------------------------
%--------------------------------------------------------
\section{Results and Discussion}
\label{sec:results}

%--------------------------------------------------------------------
\subsection{The intrinsic shape of bars in the CALIFA sample}

\begin{figure}
\includegraphics[angle=90,width=0.45\textwidth]{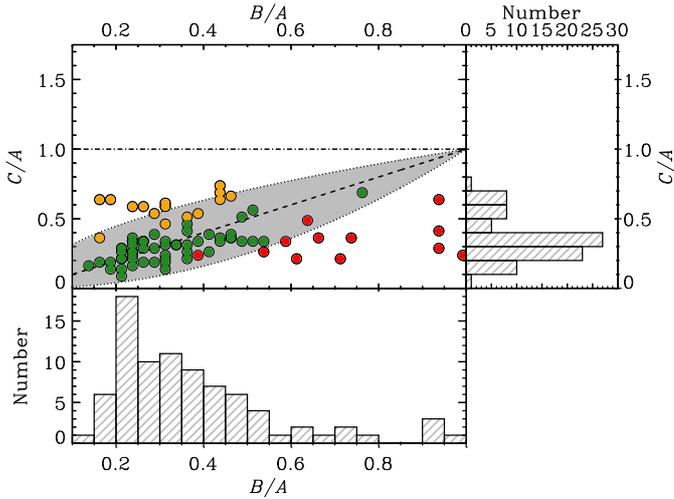}
\vspace{0.1cm}
\caption{Intrinsic axial  ratios our bar sample.   Regions are defined
  as in Fig.~\ref{fig:example3D}. Different colors represent bars with
  intrinsic  shapes as  derived  by integrating  their 2D  probability
  distribution   function   (see  text):   oblate-triaxial   off-plane
  (orange),  oblate-triaxial  in-plane   (red),  and  prolate-triaxial
  in-plane (green).  Upper  right and lower panels:  histograms of the
  intrinsic $C/A$ and $B/A$ axial ratios, respectively.  }
\label{fig:intrinsic_total}
\end{figure}

\begin{figure}
\includegraphics[width=0.45\textwidth]{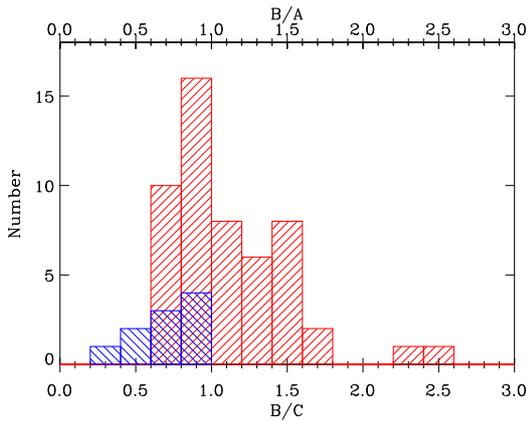}
\vspace{0.3cm}
\caption{Distribution of intrinsic semiaxis ratios $B/C$ and $B/A$ for
  our  sample of  prolate-triaxial  (red)  and oblate-triaxial  (blue)
  bars, respectively.   Prolate-triaxial bars close to  $B/C=1$ can be
  considered prolate axisymmetric bars.  Oblate-triaxial bars close to
  $B/A=1$ can be considered oblate axisymmetric bars.}
\label{fig:obl_pro}
\end{figure}

Figure~\ref{fig:intrinsic_total}   shows  the   distribution  of   the
intrinsic  axial  ratios   derived  for  our  sample   of  bars.   The
uncertainties in  each galaxy are omitted  for the sake of  clarity in
the figure, but they can  be found in Table~\ref{tab:results}. We
  defined four different regions in  this diagram.  They represent the
  expected  position   for  bars  with  different   intrinsic  shapes.
  Following  \citet{franx91}  we  assume  that  the  three-dimensional
  galaxy  density is  structured  as a  set  of coaligned  ellipsoids.
  Therefore,   general   cases   include   the   oblate-triaxial   and
  prolate-triaxial ellipsoids.  Since we  do not impose any limitation
  on the  relative size of the  ellipsoid semiaxes $A$, $B$,  and $C$,
  they can be  defined in-plane (when they are  flattened with respect
  to the disc equatorial plane) and off-plane (when they are elongated
  along  the polar  axis).  Special  cases of  this model  include the
  oblate axisymmetric ($B$ = $A$) and prolate axisymmetric ($C$ = $B$)
  spheroids.    Then,   we   classify   bars   in   four   categories:
  oblate-triaxial  (or  axisymmetric)  ellipsoids  in-plane  ($C/B$  <
  $B/A$)  or   off-plane  ($C/A$   >  $B/C$  and   $C/A$  <   1),  and
  prolate-triaxial  (or  axisymmetric)  ellipsoids in-plane  ($C/A$  <
  $B/C$ and $C/B$ > $B/A$) or elongated along the polar axis off-plane
  ($C/A$>1).  Figure~\ref{fig:intrinsic_total} shows the values of the
  bar intrinsic semiaxis ratios, $B/A$ and $C/A$, as obtained from the
  peak of their  2D probability distribution function  (yellow star in
  Fig.~\ref{fig:example3D}).   However,  the final  classification  for
  each bar  is performed by integrating  the 2D PDF of  semiaxes ratio
  within each  region of  the diagram  and finding  the region  with a
  higher probability to  be occupied by the bar.   Different colors in
  Fig.~\ref{fig:intrinsic_total}  represent  the  results of  our  bar
  classification.   The  probabilities  for  a given  bar  to  have  a
  particular shape are given in Table~\ref{tab:results}. We found that
  68\% and 14\% of our bars  can be classified as prolate-triaxial (or
  axisymmetric)   in-plane  and   oblate-triaxial  (or   axisymmetric)
  in-plane,    respectively.     Fig.~\ref{fig:obl_pro}    show    the
  distribution of  semiaxis ratios $B/C$  and $B/A$ for our  sample of
  prolate-triaxial   in-plane  and   oblate-triaxial  in-plane   bars,
  respectively.   Prolate-triaxial  bars  close   to  $B/C=1$  can  be
  considered as axisymmetric prolate spheroids whereas oblate-triaxial
  bars close  to $B/A=1$ can  be considered axisymmetric  oblate bars.
  Therefore, our analysis reveals that most  of the bars in the nearby
  Universe are, to first  order, prolate-triaxial (or axisymmetric) in
  the  plane of  the disc  with different  degrees of  flattening.  In
  addition, we found  that bars span a wide range  of both $B/A$ $\in$
  [0.1,1] and  $C/A$ $\in$ [0.1,0.8]  values, with median values  of $
  B/A \sim $ 0.31 and $C/A \sim $ 0.34.  

%--------------------------------------------------------
%--------------------------------------------------------

%--------------------------------------------------------
%--------------------------------------------------------

We   also  found   a  subclass   of  oblate-triaxial   bars  off-plane
corresponding  to 18\%  of our  sample. It  is worth  noting that  our
methodology does not impose a given axial ratio trend such as $A>B>C$,
but they are  allowed to be free.  This particularity  of our approach
allowed us,  for instance, to detect  and quantify the polar  bulge of
NGC~4698 \citep{corsini12}.    These galaxies present  the largest
  uncertainties in the $C/A$ semiaxis among the full sample due to the
  galaxy/bar  orientation with  respect to  the observer.   Therefore,
  their  measurements   are  also   compatible  with  a   less  exotic
  configuration   in    the   galactic   plane   (see    NGC~0036   in
  Fig.~\ref{fig:example3D} for  an example). Other  interesting cases
in our sample  are the four bars with a  flattened oblate axisymmetric
structure,  namely UGC~02134,  NGC~4185, NGC~5520,  and NGC~5602.   We
speculate  that   these  might  be   examples  of  lenses   that  were
photometrically fitted as normal bars.

The  intrinsic shape  of  galactic bars  has  been rarely  constrained
observationally.  \citet{kormendy82} suggested that bars are generally
triaxial ellipsoids and  used an isophotal analysis  to estimate their
typical  semiaxes ratios  to be  $B/A  = 0.2$  and $C/A  = 0.1$,  with
extreme   cases   such   as   $B/A   <  0.37$   and   $0.07<   C/A   <
0.25$. \citet{aguerri01b}  used the photometric approach  described in
\citet{varela92}  to  derive  the  intrinsic   shape  of  the  bar  in
NGC~5850. They  found a best  solution with a triaxial  ellipsoid with
$0.8  <   B/A  <  1$   and  $0.3  <   C/A  <  0.4$.    More  recently,
\citet{compere14} used a new  3D photometric decomposition approach to
derive  the  intrinsic  shape  of  a sample  of  six  bars  in  nearby
galaxies. Similarly to us, they found  that the $C/A$ semiaxis is more
difficult to constrain due to projection problems, and that 5 out of 6
bars   are  compatible   with  our   definition  of   prolate-triaxial
ellipsoids, with the remaining  being oblate-triaxial off-plane.  They
obtain mean values of $B/A = 0.24$ and $C/A = 0.31$ in reasonable good
agreement with our findings.

The analysis of the projected,  or deprojected, values of the observed
bar  ellipticity are  more common  in  the literature  since they  are
generally       associated      with       the      bar       strength
\citep{abrahammerrifield00}.  \citet{martin95} used  a 2D deprojection
method considering  the bar to  be infinitesimally thin to  derive the
deprojected values of the ellipticity.   He found that bars cover from
$0.2 < b/a < 1$ with a  $\langle b/a \rangle = 0.49$. A similar result
($\langle   b/a    \rangle   =   0.48$)   was    recently   found   by
\citet{diazgarcia16} using  a sample of  barred galaxies from  the S4G
survey  \citep{sheth10} and  the deprojection  technique developed  by
\citet{gadotti07}. Both  results show slightly larger  values of $B/A$
than those found  in this paper.  The uncertainties  in the derivation
of the deprojected  values of the bar axis ratio  $b/a$ were estimated
using mock galaxies in \citet{zou14}. They found that considering bars
as 2D thin  ellipsoids, the deprojected ellipticity of the  bar can be
recovered within a 10\% error.

Our analysis  of the intrinsic  shape of  bars suggests that  they are
generally  prolate-triaxial ellipsoids  with  an intrinsic  flattening
($C/A  \sim   0.3$),  which  corresponds  to   the  typical  intrinsic
flattening   of    galactic   discs   at   this    range   of   masses
\citep{sanchezjanssen10,   rodriguezpadilla13}.   This   is  in   good
agreement with the idea that bars form out of disc material, therefore
keeping  their  vertical   shape  \citep{sparkesellwood87}.   However,
numerical  simulations  have  shown  how  buckling  instabilities  can
quickly change this vertical structure producing a thicker boxy/peanut
shaped component  that sticks out of  the disc \citep{combessanders81,
  pfennigerfriedli91, debattista06}. Still, the peanut shape generally
does not  encompass the full  extent of the  bar, i.e., there  is some
flat      bar      outside       the      buckled      inner      part
\citep{lutticke00,athanassoula05,mendezabreu08b,erwindebattista13}, as
has     also      been     claimed      for     the      Milky     Way
\citep{martinezvalpuestagerhard11}.   We discussed extensively the
  effect of thick  B/P structures in the derived shape  of our bars in
  Sect~\ref{sec:bars3D} and Appendix~\ref{sec:barlens}.  We found that
  not accounting for  the presence of these structures  (as it happens
  in  $\sim$  43\%  of  our  galaxies)  produces  the  net  effect  of
  overestimating $B/A$ and $C/A$.  A rough correction assuming all our
  galaxies host  a boxy/peanut  would produce median  values of  $ B/A
  \sim  $ 0.27  and $C/A  \sim  $ 0.30,  i.e., still  prolate-triaxial
  ellipsoids with flattening even closer to that of stellar discs.

 The results described in this section provide new constraints for
  future detailed orbital analyses on barred potentials. Historically,
  2D  analysis used  a thin  prolate  potential to  describe the  bars
  \citep{papayannopoulospetrou83,athanassoula83,
    kaufmannpatsis05,sellwoodwilkinson93}.   With  the  advent  of  3D
  models  (somehow  promoted  by  the observational  evidence  of  B/P
  structures) triaxial potentials with typical axes ratios $B/A =0.25$
  and  $C/A  =0.1$,  similar  to those  observationally  found  by  by
  \citet{kormendy82},  became   common  \citep{pfenniger84,  skokos02,
    patsis03}.  However, our results suggest that bars generally have
larger values of both $B/A$ and $C/A$.  Using 3D $N$-body simulations,
\citet{pfennigerfriedli91} found that soon after the bar is formed, it
has a thin structure with $B/A \sim  0.42$ and $C/A \sim 0.33$, with a
later thickening  up to $B/A \sim  0.51$ and $C/A \sim  0.40$ when the
evolution seems to have ceased.  This result is in good agreement with
our findings and suggests that  some of the orbital analysis potential
might be revisited.

Hydrodynamical simulations have also used triaxial ellipsoids to model
the  bar potential  in 3D.   \citet{hunter88} found  a best  model for
NGC~3992  with a  bar potential  with semiaxis  ratios $B/A=0.54$  and
$C/A=0.33$ and  \citet{england89} shows that NGC~1300  can be modelled
using  a potential  with  a prolate  bar  with $B/A=C/A=0.34$.   These
results are in agreement with the semiaxis ratios derived for the bars
in our sample.

%---------------------------------------------------------------------
\subsection{The relation of the intrinsic shape of bars with galaxy properties}

\begin{figure}
\includegraphics[angle=90,width=0.45\textwidth]{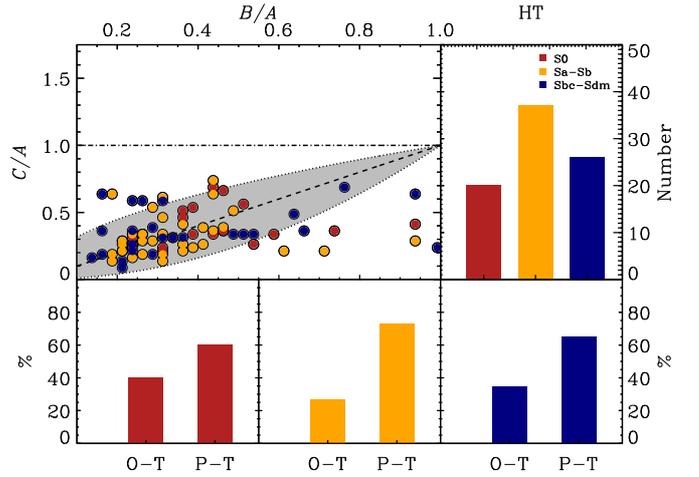}
\caption{{\it Top  left panel}: Intrinsic  axial ratios of  our CALIFA
  sample of bars  as a function of the Hubble  type (S0s, red symbols;
  Sa-Sb, yellow  symbols; Sbc-Sdm,  blue symbols).   Different regions
  represent      different      intrinsic       shapes      as      in
  Fig.~\ref{fig:example3D}.  {\it Top  right  panel}: Distribution  of
  Hubble types in our CALIFA sample (S0s: red histogram; Sa-Sb: yellow
  histogram; Sbc-Sdm: blue histogram).  Bottom panels: Distribution of
  the intrinsic shape of our CALIFA bars (O-T: oblate-triaxial or P-T:
  prolate-triaxial)  as a  function  of their  Hubble  type (S0s:  red
  histograms; Sa-Sb: yellow histograms; Sbc-Sdm: blue histograms).  }
\label{fig:intrinsic_HT}
\end{figure}

Figure~\ref{fig:intrinsic_HT}  shows  the   distribution  of  the  bar
intrinsic shapes  with the Hubble  type of the host  galaxies obtained
from \citet{walcher14}.   We found a  slight trend in the  fraction of
both  oblate-triaxial  and  prolate-triaxial bars  with  Hubble  type.
Oblate-triaxial  bars represent  40$_{-10}^{+11}$\%, 27$_{-6}^{+8}$\%,
and   34$_{-8}^{+11}$\%   of   S0,  Sa-Sb,   and   Sbc-Sdm   galaxies,
respectively.   On the  other hand,  the fraction  of prolate-triaxial
bars is  larger in spirals  than S0 galaxies  with 60$_{-11}^{+10}$\%,
73$_{-8}^{+6}$\%, and 66$_{-8}^{+6}$\%  of the bars in  S0, Sa-Sb, and
Sbc-Sdm  galaxies, respectively.   Uncertainties  were computed  using
binomial  68\%  confidence  intervals.   Figure~\ref{fig:intrinsic_HT}
also shows  a slight trend towards  bars in S0 galaxies  having larger
values of  $B/A$ and  $C/A$ than spirals.   After discarding  the four
oblate axisymmetric bars  discussed in the previous  section, we found
mean values for  bars in S0s of $\langle B/A  \rangle$ = 0.36$\pm$0.07
and $\langle C/A \rangle$ =  0.34$\pm$0.07 whereas for bars in spirals
bar we derived $\langle B/A  \rangle$ = 0.30$\pm$0.10 and $\langle C/A
\rangle$ = 0.28$\pm$0.08.

As suggested  in the previous section,  the flattening of our  bars is
similar to that expected for  discs of similar masses.  The dependence
of intrinsic semiaxis  ratios with the Hubble type  is also supportive
of  this scenario,  adding the  fact  that bars  formed in  early-type
galaxies (S0) or spirals  (Sa-Sdm), which statistically have different
intrinsic  thickening \citep{ryden06},  also  follow  the same  trend.
This result suggests  a strong relation between the  flattening of the
disc and that of the outer thin bar formed out of it.

%--------------------------------------------------------
%--------------------------------------------------------

\begin{figure}
\includegraphics[angle=90,width=0.49\textwidth]{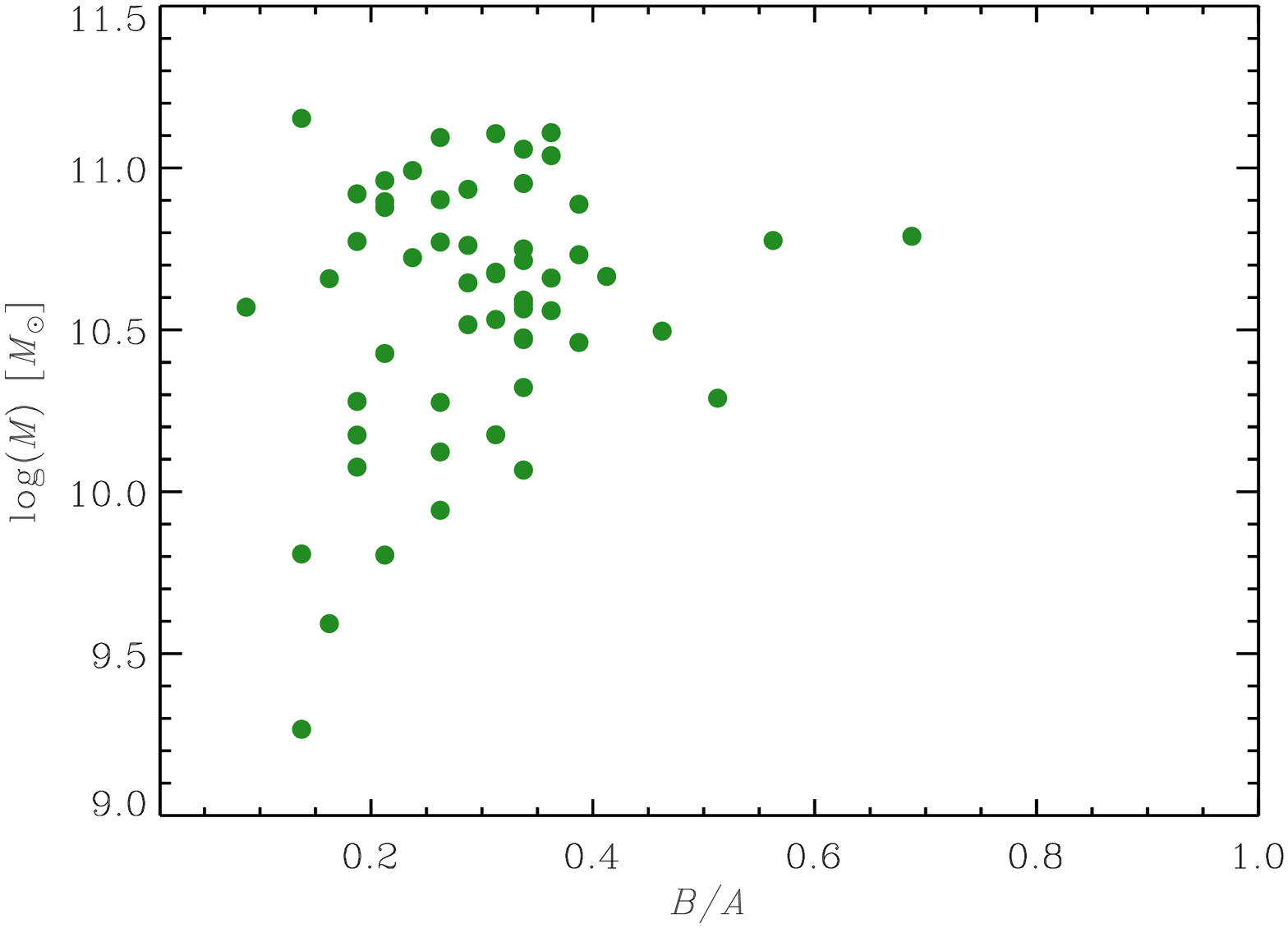}
\includegraphics[angle=90,width=0.49\textwidth]{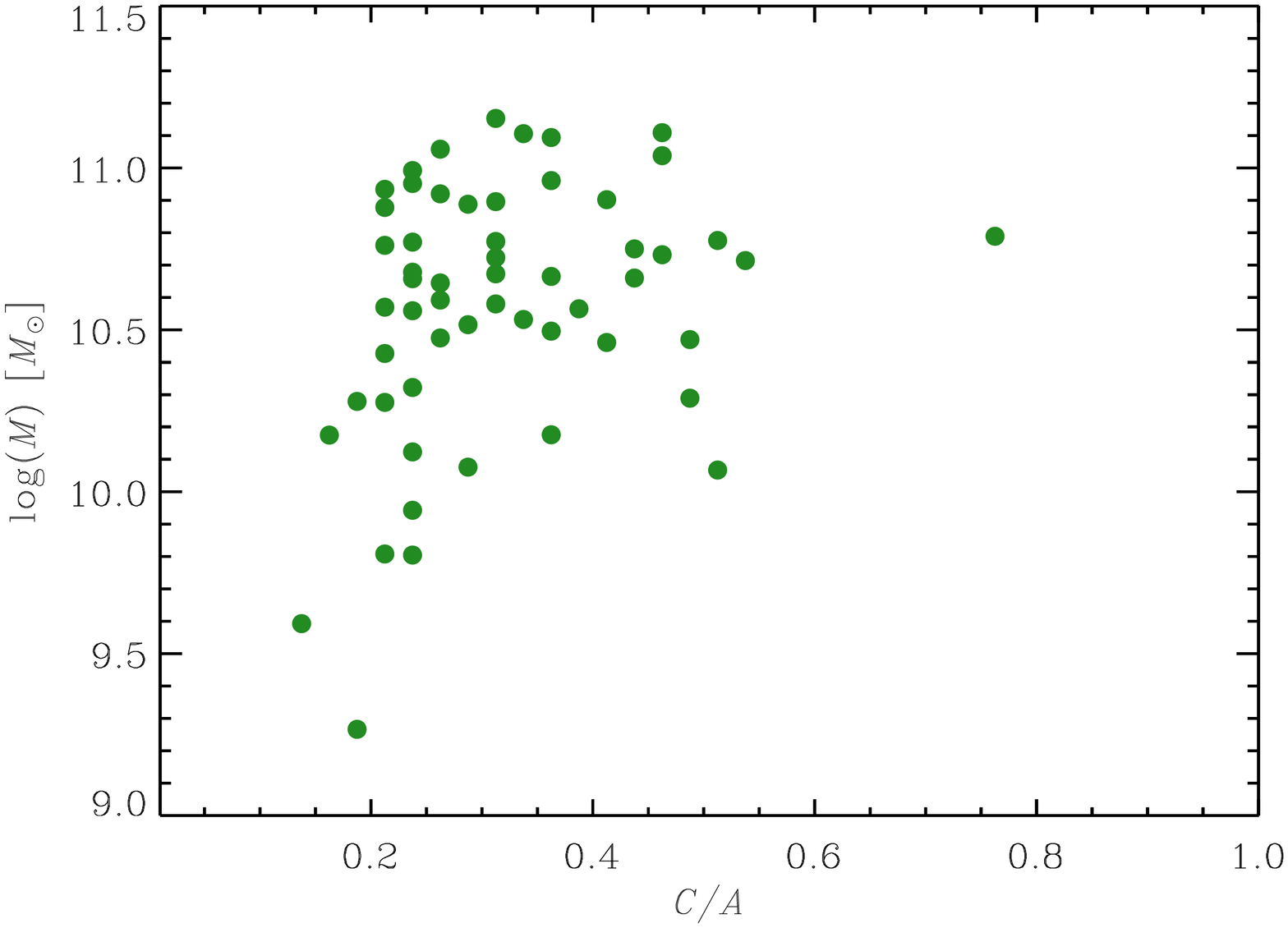}
\vspace{0.3cm}
\caption{Distribution of the axial ratio $B/A$ (upper panel) and $C/A$
  (bottom panel)  as a  function of  the galaxy  stellar mass  for our
  sample of  prolate bars.  }
\label{fig:scatter_mass}
\end{figure}

Figure~\ref{fig:scatter_mass}  shows  the  distribution of  $B/A$  and
$C/A$ for  our sample of  prolate-triaxial bars  as a function  of the
galaxy stellar mass.  A trend is  present such as bars in less massive
galaxies  are  intrinsically  more  flattened (less  $C/A$)  and  more
elliptical (less $B/A$)  than in high mass systems. In  fact, it seems
that high-mass  galaxies ($ \langle log(M_{\star}/M_{\sun})  \rangle >
$10.5) can have large values of  both $B/A$ and $C/A$ whereas low-mass
galaxies  ($  \langle  log(M_{\star}/M_{\sun}) \rangle  <  $10.5)  are
confined to low  values of $B/A$ and $C/A$.   The Spearman correlation
coefficients and p-values  are $\rho$=0.3 ($p$ =  0.02) and $\rho$=0.2
($p$=0.2) for  $B/A$ and  $C/A$, respectively.  Thus,  the correlation
with $C/A$  cannot be  considered as statistically  significative.  In
\citet{mendezabreu17} we  discussed the  relation between  Hubble type
and  galaxy  mass  for  the  CALIFA sample,  with  later  types  being
systematically less  massive galaxies.  We  found a similar,  but less
strong  trend for  our  limited sample  of 83  barred  galaxies (S0  $
\langle  log(M_{\star}/M_{\sun})  \rangle  =  $10.7$\pm$0.2;  Sa-Sb  $
\langle  log(M_{\star}/M_{\sun}) \rangle  =  $10.6$\pm$0.3; Sbc-Sdm  $
\langle    log(M_{\star}/M_{\sun})     \rangle    =    $10.1$\pm$0.6).
Table~\ref{tab:intrinsic}  shows  the  mean values  of  the  intrinsic
flattening of  our sample of prolate  bars with both stellar  mass and
Hubble type.  The  relation with the Hubble type is  still present for
the low mass bin, but it is not clear for the high mass one.  However,
the  low   number  statistics  makes  difficult   to  extract  further
conclusions on  whether is  the mass  or the  Hubble type  driving our
correlations.

%-----------------------------------
\begin{table}%[!t]
\caption{Mean values of the intrinsic  semiaxes ratios $B/A$ and $C/A$
  of  our sample  of prolate  bars as  a function  of mass  and Hubble
  type.}
\begin{center}
\begin{tabular}{|c|c|c|c|}
\hline
                       &  S0             & Sa-Sb             &  Sbc-Sdm                \\
\hline
$log(M_{\star}/M_{\sun}) < 10.5$; $B/A$ & 0.31$\pm$0.07   & 0.28$\pm$0.12     &  0.29$\pm$0.13\\ 
$log(M_{\star}/M_{\sun}) < 10.5$; $C/A$ & 0.40$\pm$0.09   & 0.29$\pm$0.13     &  0.24$\pm$0.07\\

$log(M_{\star}/M_{\sun}) > 10.5$; $B/A$ & 0.37$\pm$0.10   & 0.30$\pm$0.08     &  0.38$\pm$0.20\\
$log(M_{\star}/M_{\sun}) > 10.5$; $C/A$ & 0.33$\pm$0.10   & 0.28$\pm$0.08     &  0.36$\pm$0.18\\
    \hline
\end{tabular}
\end{center}
\label{tab:intrinsic}
\end{table}
%-------------------------------------

%--------------------------------------------------------
%--------------------------------------------------------

\begin{figure*}
\includegraphics[bb=54 94 558 738,angle=90,width=0.4\textwidth]{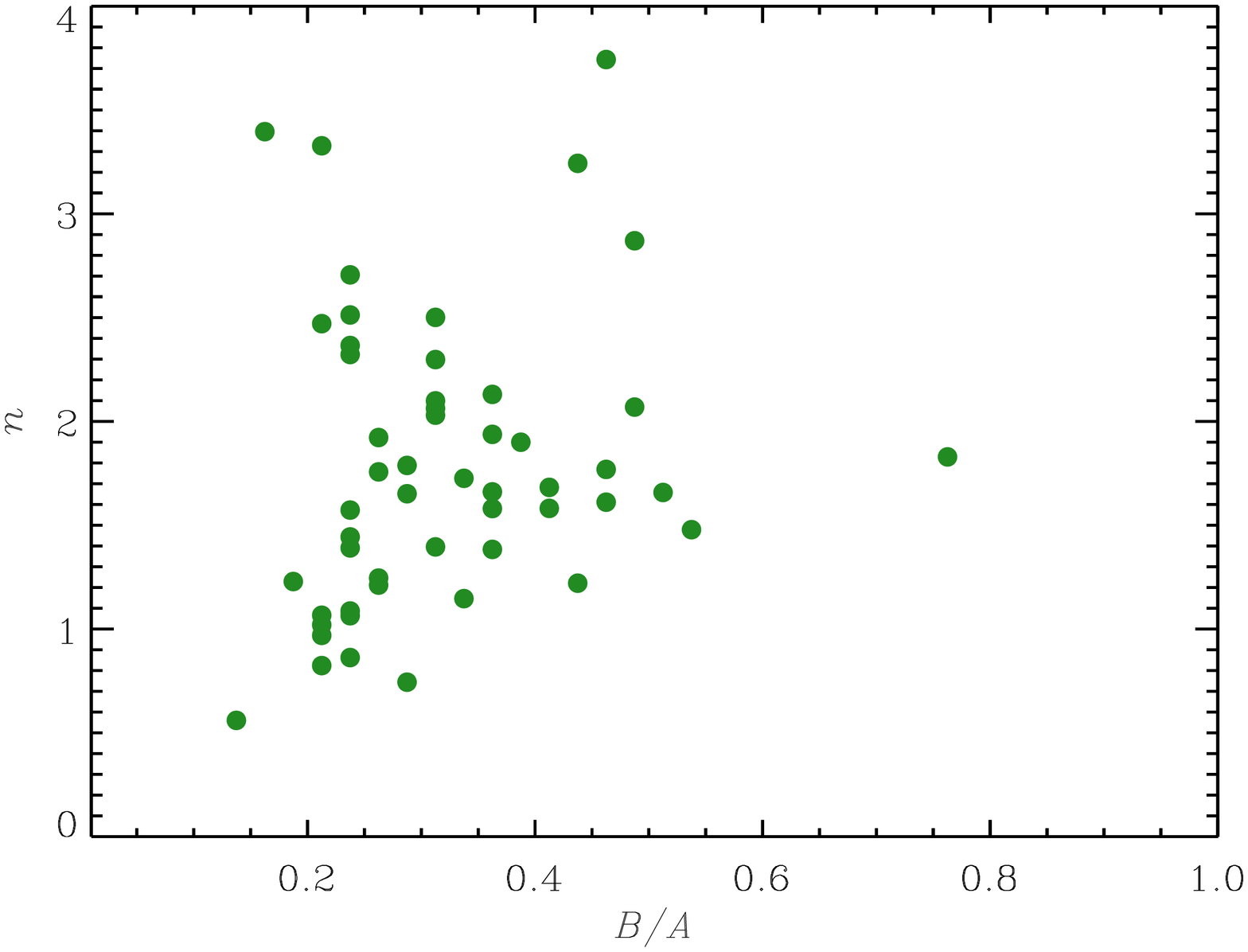}
\includegraphics[bb=54 94 558 738,angle=90,width=0.4\textwidth]{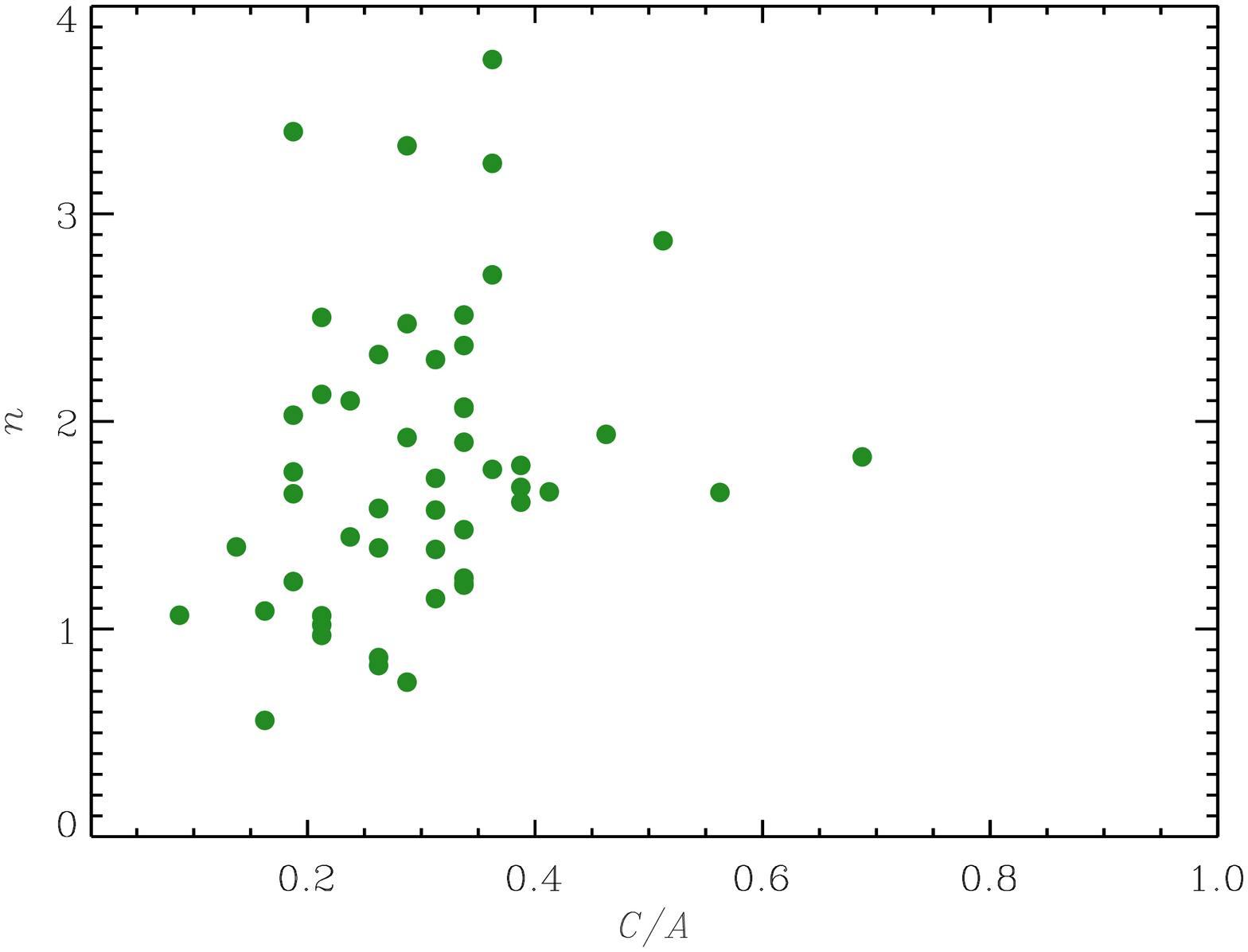}
\includegraphics[bb=54 94 558 738,angle=90,width=0.4\textwidth]{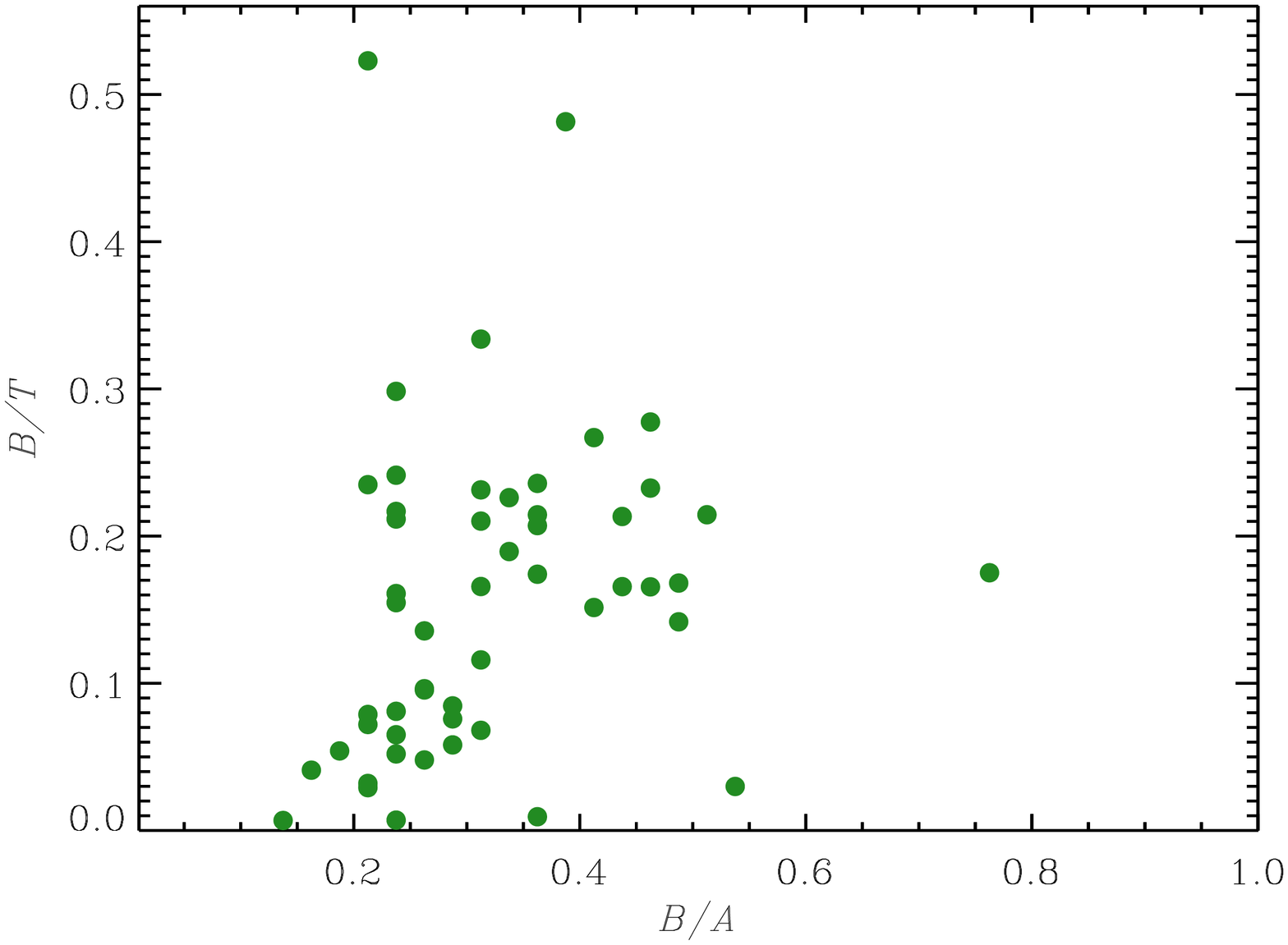}
\includegraphics[bb=54 94 558 738,angle=90,width=0.4\textwidth]{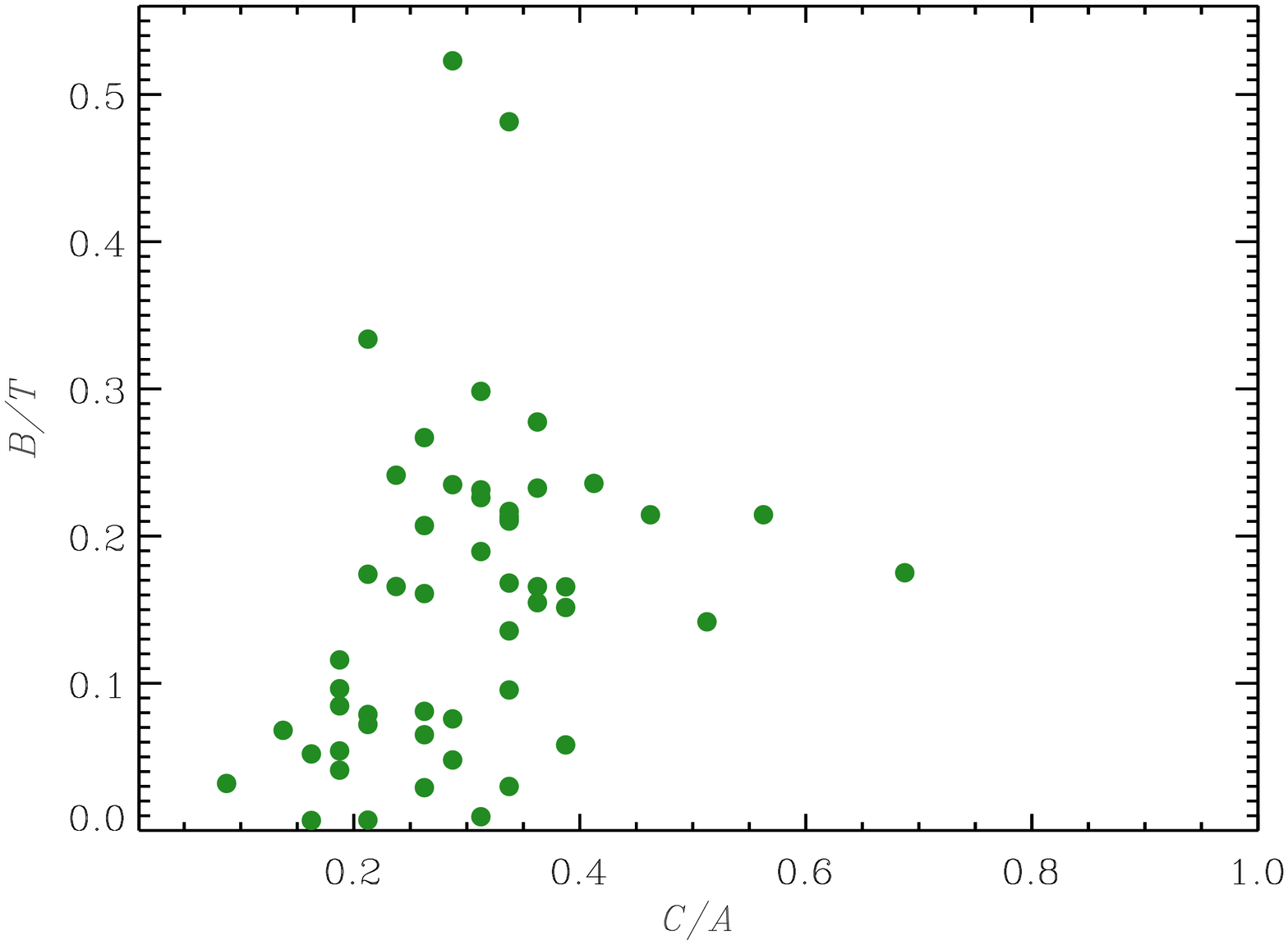}
\includegraphics[bb=54 94 558 738,angle=90,width=0.4\textwidth]{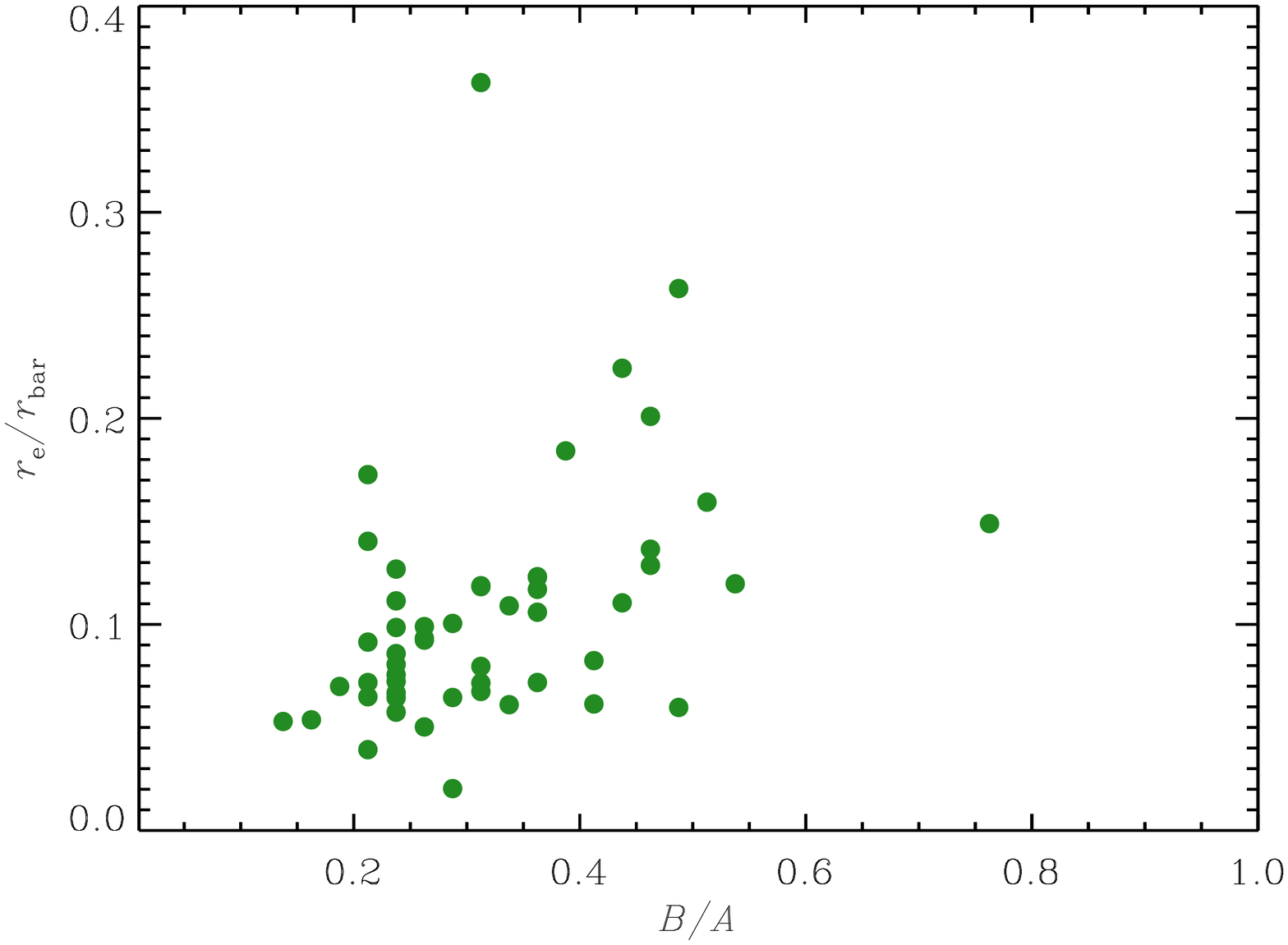}
\includegraphics[bb=54 94 558 738,angle=90,width=0.4\textwidth]{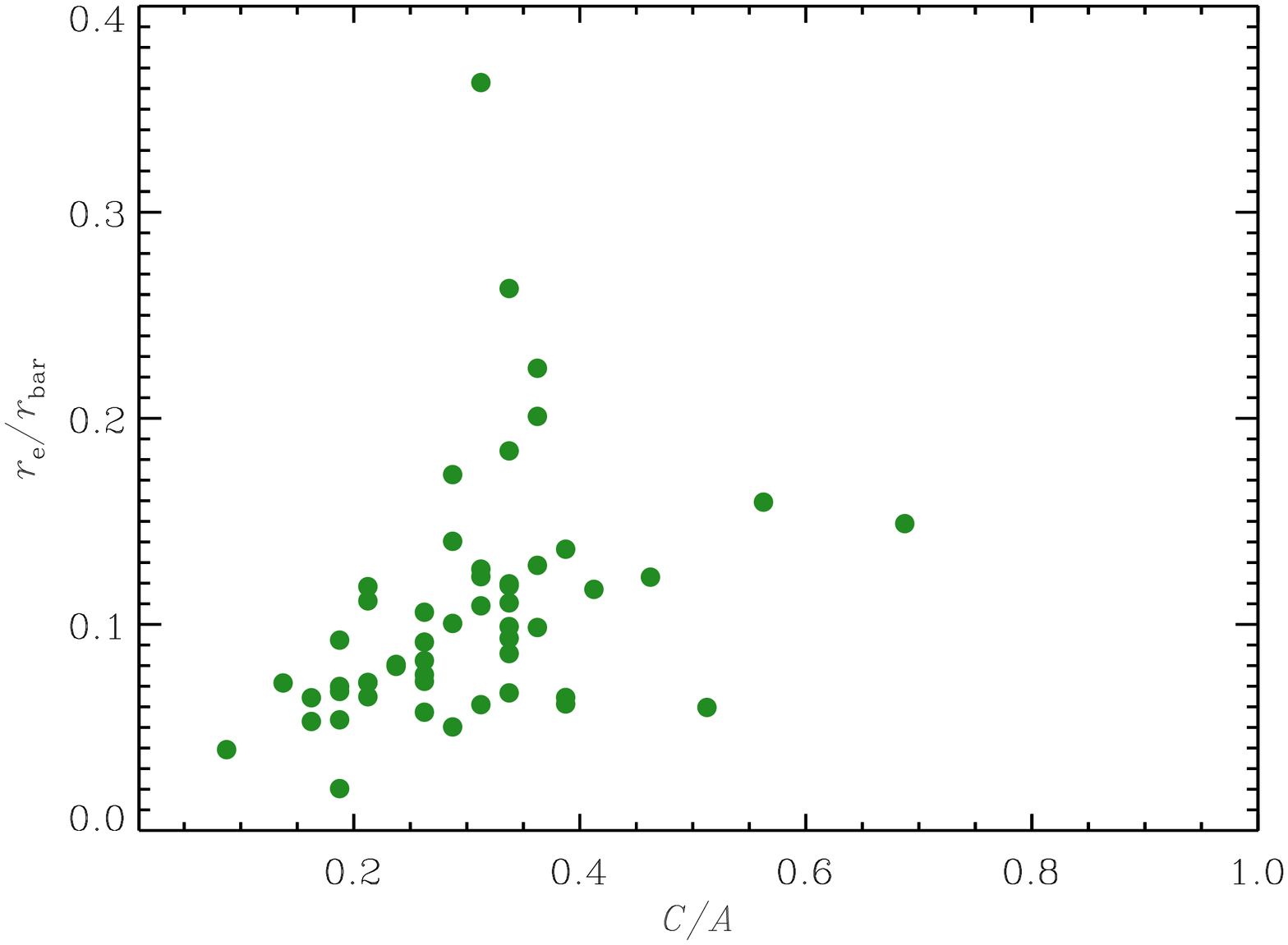}
\includegraphics[bb=54 94 558 738,angle=90,width=0.4\textwidth]{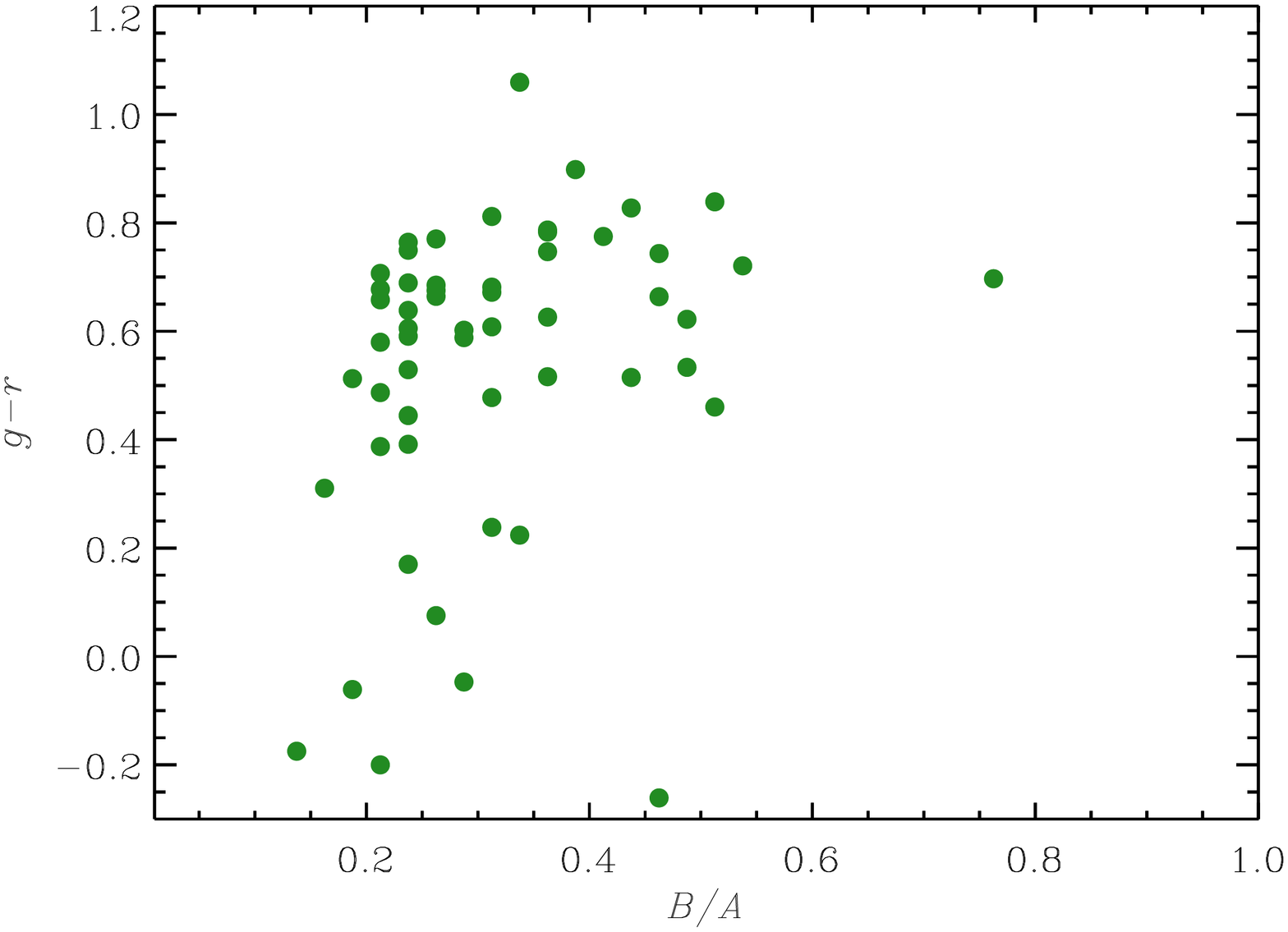}
\includegraphics[bb=54 94 558 738,angle=90,width=0.4\textwidth]{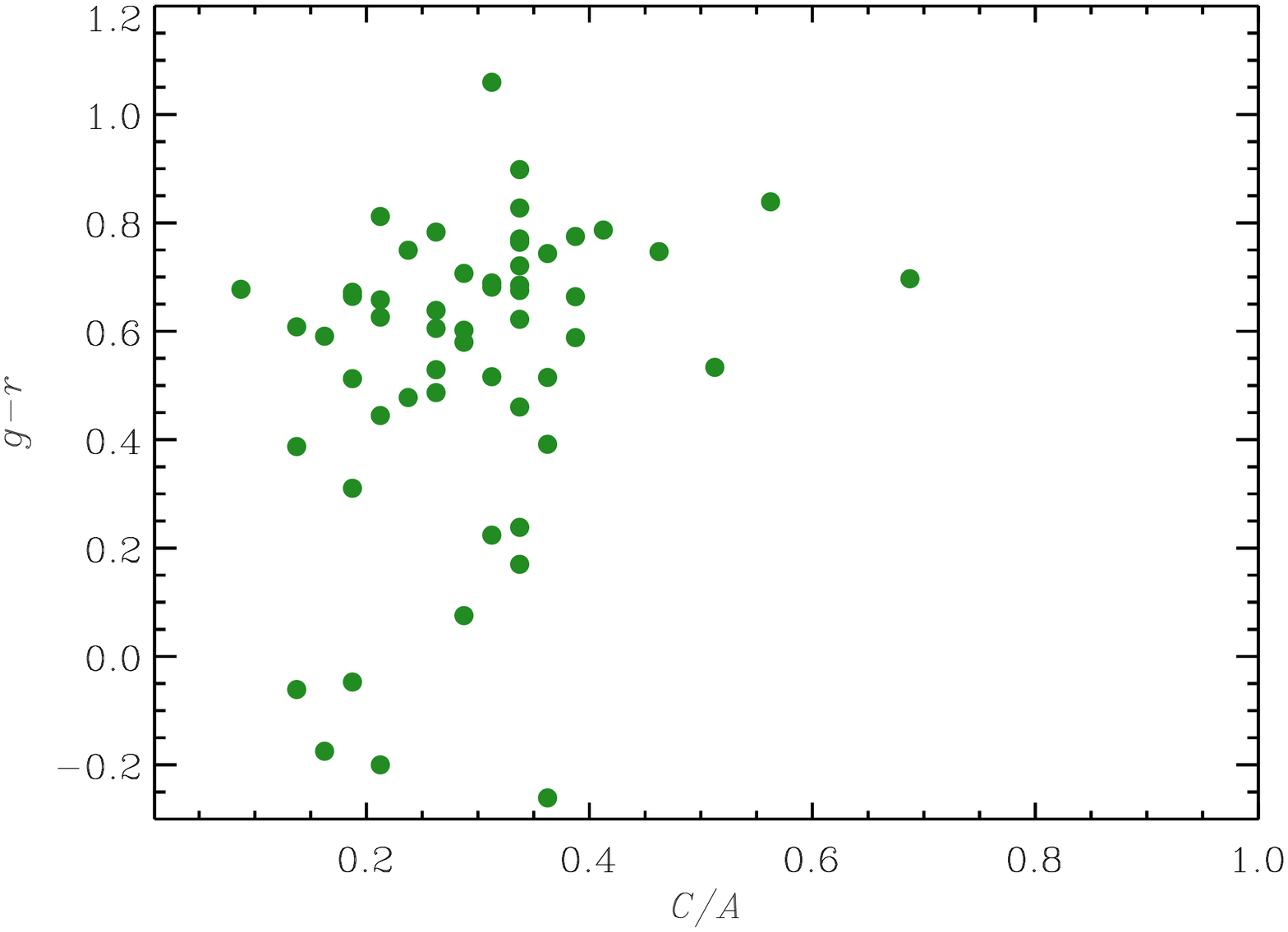}
\caption{Distribution of the intrinsic axis ratios $B/A$ (left panels)
  and  $C/A$ (right  panels)  for  our sample  of  prolate  bars as  a
  function of   (from  top to bottom)  bulge S\'ersic  index ($n$),
  bulge-to-total luminosity ratio ($B/T$),  the bulge effective radius
  over bar  radius ratio ($r_{\rm  e}/r_{\rm bar}$), and the  color of
  the disc ($g-r$).  }
\label{fig:scatter_corr}
\end{figure*}

We  also look  for correlations  between  the intrinsic  shape of  our
prolate-triaxial bars with the properties of the individual structures
shaping our  galaxies, i.e., bulges,  discs, and the  bars themselves.
Fig.~\ref{fig:scatter_corr}     shows     the     most     significant
trends. Regarding the bulge properties, we found that either $B/A$ and
$C/A$ correlates with the S\'ersic index and $B/T$ ratio. The relation
is stronger  with the $B/T$ ratio  (Fig.~\ref{fig:scatter_corr}) where
we found  a Spearman  correlation coefficient  $\rho$=0.35 ($p$=0.009)
and $\rho$=0.41 ($p$=0.002) for  $B/A$ and $C/A$, respectively.  These
correlations  indicate that  more  prominent (larger  $B/T$) and  more
concentrated  (larger $n$)  bulges  are related  with thicker  (larger
$C/A$) and  more circular bars  (larger $B/A$). Despite   the fact
  that we  did not find any  correlation with the effective  radii of
the  bulges ($r_{\rm  e}$),  we  did find  a  strong correlation  when
normalised by the bar radius  ($r_{\rm e}/r_{\rm bar}$).  Our analysis
shows  $\rho$=0.47 ($p$=0.0004)  and  $\rho$=0.52  ($p$=8e-5) for  the
relation  between  $r_{\rm  e}/r_{\rm  bar}$  vs.   $B/A$  and  $C/A$,
respectively   (see   Fig.~\ref{fig:scatter_corr}).   Regarding   disc
parameters, we  only found  a weak correlation  of the  bars intrinsic
shape with  the $g-r$  color of the  disc. Fig.~\ref{fig:scatter_corr}
shows  these  results   where  we  find  a   Spearman  coefficient  of
$\rho$=0.34 ($p$=0.02) and $\rho$=0.25 ($p$=0.08) for $B/A$ and $C/A$,
respectively. Therefore,  redder discs have thicker  and more circular
bars (but  notice that  the relation with  $C/A$ is  not statistically
significant).  Finally, we explored  the photometric properties of the
bars.  We  found  that  $C/A$  correlates  with  all  bar  parameters:
$\mu_{\rm 0, bar}$ ($\rho$= -0.46, $p$=0.0003), $r_{\rm bar}$ ($\rho$=
-0.33, $p$=0.01), and $Bar/T$ ratio  ($\rho$= -0.32, $p$=0.02). On the
other hand,  $B/A$ did  not correlate with  any bar  parameter.  These
relations  point towards  more  prominent  (larger $Bar/T$),  brighter
(larger $\mu_{\rm 0, bar}$), and  shorter (smaller $r_{\rm bar}$) bars
being thicker (larger $C/A$).

%--------------------------------------------------------
%--------------------------------------------------------

The presence of  correlations between the intrinsic shape  of the bars
and the properties of the different structures composing disc galaxies
is not surprising since bars are formed by internal secular mechanisms
out       of        disc       material       \citep{hohl71,kalnajs72,
  athanassoulamisiriotis02}, and  their evolution is  strongly related
to             the             bulge-disc-halo             interaction
\citep{weinbergkatz07,debattistasellwood00,            athanassoula13,
  long14}. Some observational studies  found that bulges influence the
size,        strength,       and        incidence       of        bars
\citep{laurikainen09,aguerri09,cheung13},   since  bulges   contribute
significantly to the radial force in discs, bulge masses and sizes can
affect bar  formation \citep{hohl76,  efstathiou82}. However,  most of
these earlier studies  considered the spheroidal component  as a halo,
not as a bulge, and more  recent studies focus mainly in the formation
of      the     bar      and     not      in     their      properties
\citep{scannapiecoathanassoula12,    kataria18}.    Therefore,    more
numerical  simulations focused  on the  bar properties  are needed  to
understand the  dependence of the  bar intrinsic shape with  the bulge
properties that we find in this paper.

%--------------------------------------------------------
\subsection{Bars vs bulges: the connection with bulge growth}

\begin{figure}
\includegraphics[angle=90,width=0.49\textwidth]{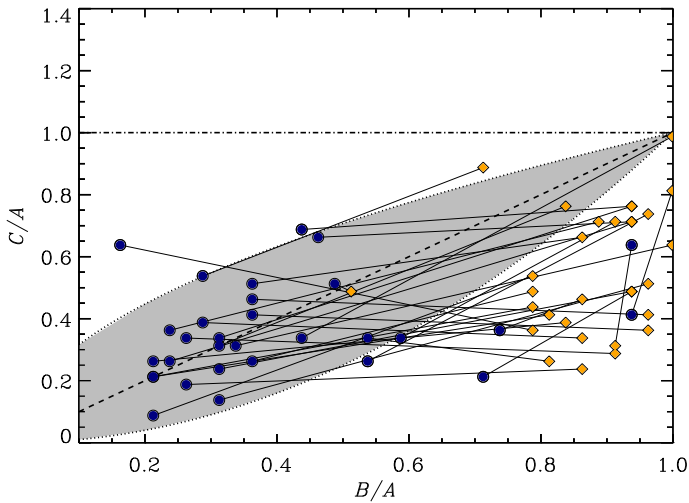}
\caption{Intrinsic axial  ratios of  our CALIFA  sample of  bars (blue
  circles) and  bulges (yellow diamonds). Structures  belonging to the
  same galaxy are linked by a solid line.}
\label{fig:bulgesvsbars}
\end{figure}

In \citet{costantin18}  we derived the  intrinsic shape of  the bulges
present  in the  CALIFA sample  of  galaxies.  We  followed a  similar
methodology as the one used in  this paper and we found that, contrary
to galactic bars, bulges tend to be nearly oblate systems (66\%), with
a smaller fraction of prolate spheroids (19\%) and triaxial ellipsoids
(15\%).   We also  found that  those bulges  sharing potential  with a
galactic bar represented  the majority of our  triaxial bulges (75\%).
Fig.~\ref{fig:bulgesvsbars} shows  the relation between  the intrinsic
axial ratios  of both  bars and  bulges for the  31  galaxies in
common  with   \citet{costantin18}.   We  can  distinguish   two  main
behaviours  in this  diagram: i)  bulges that  are thicker  than their
surrounding bars and ii) bulges  that are thinner than the surrounding
bar. We computed the differences between the intrinsic flattening
  of  both  bulges and  bars  using  the marginal  $C/A$  distribution
  obtained from the PDF for  each structure.  Since both distributions
  are independent, for  each galaxy we randomly sampled  the $C/A$ PDF
  of the bar and bulge (see Fig.~\ref{fig:example3D} for examples) and
  computed for every pair of values  whether the bulge has a larger or
  smaller $C/A$ than  the bar. We derived that 52\%  and 13\% of our
bulges are  more vertically prominent  than their surrounding  bars at
1$\sigma$  and 3$\sigma$  levels, respectively.   Similarly, we  found
that  16\%  of our  bulges  are  more  flattened  that their  bars  at
1$\sigma$ level.
In general, those bulges with $C/A$ and $B/A$ close to one, i.e., with
a nearly spherical shape, are surrounded by a prolate thinner bar.  On
the  other hand,  oblate-triaxial (or  axisimmetric) bulges  which are
flattened  in  the vertical  direction  have  an intrinsic  flattening
similar to that of the bar.

%--------------------------------------------------------
%--------------------------------------------------------

The  intrinsic  shape  of  bulges  has  been  proposed  as  a  way  to
distinguish between bulges  formed at early stages of  the Universe by
dissipative                 processes                 \citep[classical
  bulges;][]{athanassoula05,brooks16}  and  those  formed  by  secular
processes        within       the        galaxy       \citep[disc-like
  bulges][]{kormendykennicutt04}. Recent  works by \citet{costantin18}
and  Costantin et  al.  (in  prep.)  have  shown that  oblate-triaxial
bulges which are intrinsically  flattened in-plane might be considered
as disc-like bulges whereas  oblate-triaxial non-flattened bulges show
properties of classical bulges.  The main internal mechanism to form a
disc-like bulge is  gas flowing inwards to the galaxy  center thank to
the  torque  exerted  by  a  bar  structure  \citep{pfennigernorman90,
  friedlibenz95}.  This gas  then accumulates in the  center forming a
new                    rotationally-supported                    bulge
\citep{kormendykennicutt04,boone07,wozniakmicheldansac09}.    We  have
demonstrated how the intrinsic flattening  of the majority of our long
bars  is strongly  related to  vertical  extension of  the host  disc,
therefore, the results shown in Fig.~\ref{fig:bulgesvsbars}, with 16\%
of the bulges having lower intrinsic flattening than the corresponding
bar  (and also  similar to  the disc)  make these  bulges the  perfect
candidates  to be  disc-like  bulges.  On  the  other hand,  spherical
bulges hosted within thinner bars  (and thinner discs) might have been
created by  different (not  internal secular) processes  and therefore
they could represent  a population of classical  bulges.  The presence
of classical bulges, and their  coexistence with both disc-like bulges
and boxy/peanuts  structures, in barred  galaxies has been  studied in
the    literature    \citep{delorenzocaceres12,mendezabreu14,erwin15}.
However,   their  identification   generally  requires   the  use   of
photometric and  spectroscopic information. The results  of this paper
reinforce the  idea that  the intrinsic shape  of bulges  provides key
information  on their  classification,  specially  in barred  galaxies
where it  can be directly  compared with that  of the bar  (similar to
that  of the  disc). Furthermore,  the 3D  structure of  the bulge  is
derived using only  photometric information so it could  be applied to
larger samples of galaxies.

%--------------------------------------------------------
%--------------------------------------------------------
\section{Conclusions}
\label{sec:conclusions}

We present  the first statistical study  on the 3D intrinsic  shape of
galactic  bars in  the nearby  Universe. We  analysed a  sample of  83
barred galaxies extracted from the  CALIFA-DR3 survey with the galaXYZ
code \citep{mendezabreu10,costantin18}.   This method is  purely based
on photometric information and uses the projected geometric parameters
(ellipticities and  position angles) of  both bars and  discs obtained
from    a    careful   multi-component    photometric    decomposition
\citep{mendezabreu17}.

Our  analysis consider  galactic bars  as single  triaxial ellipsoidal
structures.  We  discuss that  our results  are representative  of the
outer  thin bar  due  to  the way  photometric  decompositions of  the
galaxies were performed. Using mock  galaxy simulations we derive that
the  possible presence  of a  vertically thick  components in  the bar
inner regions (boxy/peanut structure)  would produce an overestimation
of 0.04 in  both $B/A$ and $C/A$ intrinsic semiaxes.  Our main results
on the intrinsic shape of outer thin bars are:

\begin{itemize}

\item  Bars are  mainly  prolate-triaxial  (or axisymmetric)  in-plane
  ellipsoids  (67\%),  with  another  14\% of  bars  characterised  as
  oblate-triaxial (or axisymmetric) in-plane  ellipsoids.  They have a
  typical intrinsic intrinsic flattening C/A = 0.34, which matches the
  intrinsic  flattening of  galactic discs  of similar  stellar masses
  \citep{sanchezjanssen10}.

\item  The  intrinsic  flattening  of  our  prolate-triaxial  bars  is
  dependent of the galaxy Hubble type and galaxy stellar mass.  In our
  sample, the  Hubble type is related  to the galaxy mass,  but due to
  low  number statistics  it  is difficult  to  separate which  galaxy
  property is driving  the vertical growth of the outer  bar.  We also
  found correlations  of the bar  flattening with bulge  ($n$, $B/T$),
  disc (color $g-r$), and  bar ($\mu_{0,bar}$, $r_{\rm bar}$, $Bar/T$)
  properties, suggesting  that bar  evolution is tightly  related with
  the structures of the host galaxy.

\item We  compared the relative  flattening of  bars and bulges  for a
  subsample   of  31   galaxies   in  common   with   the  sample   of
  \citet{costantin18}.   We found  that 52\%  and 16\%  of bulges  are
  either more or  less vertically extended than  their surrounding bar
  at 1 sigma level, respectively.  Assuming that (as we demonstrate in
  this paper) the intrinsic flattening  of bars and discs are similar,
  we suggest  that these  percentages might  be representative  of the
  fraction of classical and disc-like bulges in our sample.

\end{itemize}

This paper represents  our first attempt to extend  our methodology to
derive  the  intrinsic  shape  of  galactic  structure  in  individual
galaxies beyond the  properties of bulges. The study  of the evolution
of the  intrinsic shape of  bars contains important  information about
both  the  evolution  of  these  systems and  the  creation  of  inner
structures, since the axis ratio of  the bar has been recognised as an
important quantity which drives, for  instance, the amount of gas that
is   driven    to   the   galaxy    center   by   the    bar   torques
\citep{friedlibenz93}.   Further  studies  on  the  evolution  of  the
intrinsic shape of bars with both cosmic time and dynamical properties
of the  galaxies will  be presented  in forthcoming  papers, providing
further constraints for numerical simulations.

%-----------------------------------------------------
\section*{Acknowledgements}

We  thank the  anonymous referee  for her/his  many valuable  comments
which helped to improve this  paper.  JMA and JALA acknowledge support
from the Spanish  Ministerio de Economia y  Competitividad (MINECO) by
the grant AYA2013-43188-P.  AdLC acknowledges support from the Spanish
Ministry of Economy and  Competitiveness (MINECO) grant AYA2011-24728.
EMC  is supported  by Padua  University through  grants 60A02-5857/13,
60A02-5833/14, 60A02-4434/15, and CPDA133894.   This paper is based on
data from  the Calar  Alto Legacy Integral  Field Area  Survey, CALIFA
(http://califa.caha.es), funded  by the  Spanish Ministery  of Science
under    grant   ICTS-2009-10,    and    the   Centro    Astron\'omico
Hispano-Alem\'an.

Based on  observations collected  at the Centro  Astron\'omico Hispano
Alem\'an  (CAHA) at  Calar Alto,  operated jointly  by the  Max-Planck
Institut  f\"ur  Astronomie  and  the Instituto  de  Astrof\'isica  de
Andaluc\'ia (CSIC).

%%%%%%%%%%%%%%%%%%%%%%%%%%%%%%%%%%%%%%%%%%%%%%%%%%

%%%%%%%%%%%%%%%%%%%% REFERENCES %%%%%%%%%%%%%%%%%%

% The best way to enter references is to use BibTeX:

\bibliographystyle{mnras}
\bibliography{reference} % if your bibtex file is called example.bib

%%%%%%%%%%%%%%%%%%%%%%%%%%%%%%%%%%%%%%%%%%%%%%%%%%

%%%%%%%%%%%%%%%%% APPENDICES %%%%%%%%%%%%%%%%%%%%%
\onecolumn
\appendix
\section{Structural parameters of the galaxy sample}

\begin{landscape}
\begin{tiny}
\begin{center}
\begin{longtable}{p{1.2cm} c c c c c c c c c c c c c c c c c c}%p{2.0cm} p{1.3cm} p{1.3cm} p{1.cm} p{1.3cm} p{1.3cm} p{1.3cm} p{1.3cm} p{1.9cm} p{1.9cm} p{1.7cm} p{1.7cm} p{1.5cm} p{1.5cm}}
\caption{\label{tab:results}Main properties and structural parameters of our CALIFA sample of galaxies.}\\     
\hline\hline
\multicolumn{1}{c}{Galaxy} & 
\multicolumn{1}{c}{Redshift} & 
\multicolumn{1}{c}{log($M_{\star}/M_{\sun}$)}& 
\multicolumn{1}{c}{HT} &
\multicolumn{1}{c}{$g-r$ (disc)} &
\multicolumn{1}{c}{$\mu_{0}$} &
\multicolumn{1}{c}{$n$} &
\multicolumn{1}{c}{$r_{\rm e}$} &
\multicolumn{1}{c}{$q_{\rm d}$}& 
\multicolumn{1}{c}{$q_{\rm bar}$}& 
\multicolumn{1}{c}{$PA_{\rm d}$}&
\multicolumn{1}{c}{$PA_{\rm bar}$}& 
\multicolumn{1}{c}{$B/A$}&
\multicolumn{1}{c}{$C/A$}&
\multicolumn{1}{c}{\textsl{P}(pro. o)}&
\multicolumn{1}{c}{\textsl{P}(pro. i)}&
\multicolumn{1}{c}{\textsl{P}(obl. o)}&
\multicolumn{1}{c}{\textsl{P}(obl. i)}&
\multicolumn{1}{c}{V. class}\\ 
\multicolumn{1}{c}{(1)}&
\multicolumn{1}{c}{(2)}& 
\multicolumn{1}{c}{(3)}& 
\multicolumn{1}{c}{(4)}&
\multicolumn{1}{c}{(5)}& 
\multicolumn{1}{c}{(6)}&
\multicolumn{1}{c}{(7)}& 
\multicolumn{1}{c}{(8)}&
\multicolumn{1}{c}{(9)}&
\multicolumn{1}{c}{(10)}&
\multicolumn{1}{c}{(11)}&
\multicolumn{1}{c}{(12)}&
\multicolumn{1}{c}{(13)}&
\multicolumn{1}{c}{(14)}&
\multicolumn{1}{c}{(15)}&
\multicolumn{1}{c}{(16)}&
\multicolumn{1}{c}{(17)}&
\multicolumn{1}{c}{(18)}&
\multicolumn{1}{c}{(19)}\\ 
\hline
\endfirsthead
\caption{continued.}\\
\hline\hline
\multicolumn{1}{c}{Galaxy} & 
\multicolumn{1}{c}{Redshift} & 
\multicolumn{1}{c}{log($M_{\star}/M_{\sun}$)}& 
\multicolumn{1}{c}{HT} &
\multicolumn{1}{c}{$g-r$ (disc)} &
\multicolumn{1}{c}{$\mu_{0}$} &
\multicolumn{1}{c}{$n$} &
\multicolumn{1}{c}{$r_{\rm e}$} &
\multicolumn{1}{c}{$q_{\rm d}$}& 
\multicolumn{1}{c}{$q_{\rm bar}$}& 
\multicolumn{1}{c}{$PA_{\rm d}$}&
\multicolumn{1}{c}{$PA_{\rm bar}$}& 
\multicolumn{1}{c}{$B/A$}&
\multicolumn{1}{c}{$C/A$}&
\multicolumn{1}{c}{P(prol. op)}&
\multicolumn{1}{c}{P(prol. ip)}&
\multicolumn{1}{c}{P(obl. op)}&
\multicolumn{1}{c}{P(obl. ip)}&
\multicolumn{1}{c}{V. class}\\ 
\multicolumn{1}{c}{(1)}&
\multicolumn{1}{c}{(2)}& 
\multicolumn{1}{c}{(3)}& 
\multicolumn{1}{c}{(4)}&
\multicolumn{1}{c}{(5)}& 
\multicolumn{1}{c}{(6)}&
\multicolumn{1}{c}{(7)}& 
\multicolumn{1}{c}{(8)}&
\multicolumn{1}{c}{(9)}&
\multicolumn{1}{c}{(10)}&
\multicolumn{1}{c}{(11)}&
\multicolumn{1}{c}{(12)}&
\multicolumn{1}{c}{(13)}&
\multicolumn{1}{c}{(14)}&
\multicolumn{1}{c}{(15)}&
\multicolumn{1}{c}{(16)}&
\multicolumn{1}{c}{(17)}&
\multicolumn{1}{c}{(18)}&
\multicolumn{1}{c}{(19)}\\ 
\hline
\endhead
\hline
\endfoot            
UGC00036 &    0.0206  &      10.764  &      Sab   &  0.78    &     19.1   &      1.52    &    0.54   &      0.47$\pm$0.01   &  0.668$\pm$0.007 &  17.9$\pm$0.7  &  133.3$\pm$0.4 &  0.44$^{0.46}_{0.26}$ &     0.64$^{0.96}_{0.21}$ & 0.01  & 0.44 & 0.55 & 0.00  & Barlens?         \\
NGC0036 &    0.0197   &      10.746  &      Sb    &  0.69    &     20.2   &      4.93    &    2.46   &      0.57$\pm$0.01   &  0.465$\pm$0.007 &  16.7$\pm$0.7  &  124.1$\pm$0.4 &  0.31$_{0.16}^{0.31}$ &     0.62$_{0.16}^{1.04}$ & 0.07  & 0.26 & 0.66 & 0.01  & Barlens          \\
NGC0165 &    0.0192   &      10.482  &      Sb    &  0.60    &     20.3   &      0.74    &    0.78   &      0.81$\pm$0.01   &  0.295$\pm$0.007 &  83.3$\pm$0.7  &  62.6$\pm$0.4  &  0.29$_{0.14}^{0.34}$ &     0.29$_{0.11}^{0.46}$ & 0.00  & 0.66 & 0.27 & 0.07  & Barlens?         \\
NGC0180 &    0.0172   &      10.597  &      Sb    &  0.59    &     20.2   &      1.09    &    0.67   &      0.682$\pm$0.007 &  0.227$\pm$0.006 &  165.8$\pm$0.3 &  141.9$\pm$0.2 &  0.24$_{0.14}^{0.26}$ &     0.16$_{0.09}^{0.26}$ & 0.00  & 0.82 & 0.10 & 0.08  & Barlens          \\
NGC0214 &    0.0148   &      10.467  &      Sbc   &  0.60    &     19.0   &      1.10    &    0.56   &      0.705$\pm$0.007 &  0.564$\pm$0.006 &  57.4$\pm$0.3  &  60.1$\pm$0.2  &  0.64$_{0.39}^{0.84}$ &     0.49$_{0.09}^{0.69}$ & 0.00  & 0.27 & 0.09 & 0.64  & None             \\
NGC0364 &    0.0166   &      10.647  &      E7    &  0.83    &     19.9   &      1.55    &    0.58   &      0.73$\pm$0.01   &  0.548$\pm$0.007 &  33.3$\pm$0.7  &  91.9$\pm$0.4  &  0.44$_{0.24}^{0.46}$ &     0.69$_{0.16}^{1.11}$ & 0.19  & 0.32 & 0.46 & 0.03  & Barlens          \\
NGC0447 &    0.0183   &      10.901  &      Sa    &  0.76    &     21.2   &      2.37    &    1.25   &      0.858$\pm$0.007 &  0.270$\pm$0.006 &  28.2$\pm$0.3  &  17.5$\pm$0.2  &  0.24$_{0.12}^{0.31}$ &     0.34$_{0.11}^{0.49}$ & 0.00  & 0.66 & 0.30 & 0.04  & Barlens          \\
IC1683  &    0.0159   &      10.461  &      Sb    &  0.72    &     20.8   &      1.68    &    0.53   &      0.59$\pm$0.01   &  0.509$\pm$0.007 &  13.5$\pm$0.7  &  157.0$\pm$0.4 &  0.41$_{0.21}^{0.46}$ &     0.39$_{0.11}^{0.59}$ & 0.00  & 0.66 & 0.22 & 0.12  & None             \\
NGC0551 &    0.0170   &      10.528  &      Sbc   &  0.68    &     19.5   &      1.07    &    0.43   &      0.44$\pm$0.01   &  0.194$\pm$0.007 &  135.9$\pm$0.7 &  160.2$\pm$0.4 &  0.21$_{0.09}^{0.24}$ &     0.09$_{0.04}^{0.16}$ & 0.00  & 0.83 & 0.02 & 0.15  & None              \\
NGC0570 &    0.0178   &      10.879  &      Sb    &  0.75    &     20.4   &      1.86    &    1.00   &      0.90$\pm$0.01   &  0.335$\pm$0.007 &  138.0$\pm$0.7 &  94.0$\pm$0.4  &  0.29$_{0.14}^{0.34}$ &     0.54$_{0.19}^{0.94}$ & 0.02  & 0.29 & 0.68 & 0.02  & Barlens           \\      
UGC01271 &    0.0164  &      10.532  &      S0a   &  1.06    &     20.0   &      1.73    &    0.68   &      0.63$\pm$0.01   &  0.427$\pm$0.007 &  93.9$\pm$0.7  &  52.1$\pm$0.4  &  0.34$_{0.14}^{0.36}$ &     0.31$_{0.11}^{0.54}$ & 0.00  & 0.68 & 0.26 & 0.06  & Barlens            \\
NGC0716 &    0.0147   &      10.555  &      Sb    &  0.69    &     18.9   &      ---     &    ---    &      0.48$\pm$0.01   &  0.469$\pm$0.006 &  62$\pm$1.0    &  13.2$\pm$0.5  &  0.26$_{0.11}^{0.29}$ &     0.34$_{0.11}^{0.49}$ & 0.00  & 0.75 & 0.22 & 0.03  & None               \\
NGC0776 &    0.0159   &      10.597  &      Sb    &  0.74    &     20.3   &      1.36    &    0.59   &      0.90$\pm$0.01   &  0.349$\pm$0.007 &  146.4$\pm$0.7 &  130.9$\pm$0.4 &  0.31$_{0.14}^{0.36}$ &     0.46$_{0.19}^{0.76}$ & 0.00  & 0.43 & 0.53 & 0.04  & None              \\
NGC0842 &    0.0124   &      10.460  &      S0    &  0.73    &     19.8   &      2.60    &    0.57   &      0.525$\pm$0.004 &  0.540$\pm$0.003 &  145.1$\pm$0.3 &  155.6$\pm$0.2 &  0.74$_{0.46}^{0.79}$ &     0.36$_{0.14}^{0.54}$ & 0.00  & 0.28 & 0.04 & 0.68  & None              \\
UGC01918 &   0.0165   &      10.551  &      Sb    &  0.76    &     20.2   &      1.14    &    0.31   &      0.45$\pm$0.01   &  0.62$\pm$0.02   &  117.3$\pm$0.9 &  13.7$\pm$0.8  &  0.44$_{0.26}^{0.51}$ &     0.74$_{0.24}^{1.20}$ & 0.27  & 0.25 & 0.48 & 0.00  & None              \\
NGC0976 &    0.0139   &      10.774  &      Sbc   &  0.70    &     19.1   &      1.83    &    0.77   &      0.786$\pm$0.007 &  0.777$\pm$0.006 &  159.6$\pm$0.3 &  10.6$\pm$0.2  &  0.76$_{0.59}^{0.79}$ &     0.69$_{0.24}^{1.06}$ & 0.19  & 0.31 & 0.18 & 0.31  & None              \\
UGC02134 &    0.0149  &      10.479  &      Sb    &  0.74    &     19.7   &      2.61    &    0.93   &      0.441$\pm$0.009 &  0.470$\pm$0.007 &  105.5$\pm$0.5 &  103.9$\pm$0.4 &  0.94$_{0.86}^{0.96}$ &     0.29$_{0.16}^{0.41}$ & 0.00  & 0.04 & 0.00 & 0.96  & BP?            \\
NGC1093 &    0.0172   &      10.392  &      Sbc   &  0.62    &     20.0   &      2.07    &    1.70   &      0.67$\pm$0.01   &  0.448$\pm$0.007 &  96.5$\pm$0.7  &  116.5$\pm$0.4 &  0.49$_{0.29}^{0.56}$ &     0.34$_{0.09}^{0.51}$ & 0.00  & 0.59 & 0.14 & 0.27  & Barlens             \\
UGC02403 &    0.0133  &      10.422  &      Sb    &  0.86    &     20.7   &      1.04    &    0.62   &      0.46$\pm$0.01   &  0.286$\pm$0.007 &  148.4$\pm$0.7 &  140.8$\pm$0.4 &  0.39$_{0.21}^{0.64}$ &     0.24$_{0.09}^{0.29}$ & 0.00  & 0.36 & 0.01 & 0.63  & None             \\
NGC1645 &    0.0160   &      10.673  &      S0a   &  0.12    &     19.6   &      0.95    &    0.60   &      0.48$\pm$0.01   &  0.618$\pm$0.007 &  86.5$\pm$0.7  &  22.9$\pm$0.4  &  0.39$_{0.21}^{0.41}$ &     0.54$_{0.16}^{0.89}$ & 0.00  & 0.47 & 0.52 & 0.01  & Barlens              \\
NGC1666 &    0.0090   &      10.496  &      S0a   &  0.74    &     19.7   &      1.94    &    0.31   &      0.879$\pm$0.004 &  0.389$\pm$0.003 &  147.5$\pm$0.3 &  141.5$\pm$0.2 &  0.36$_{0.19}^{0.44}$ &     0.46$_{0.16}^{0.74}$ & 0.00  & 0.47 & 0.46 & 0.08  & None              \\
NGC1667 &    0.0149   &      10.620  &      Sbc   &  0.72    &     18.4   &      1.48    &    0.29   &      0.687$\pm$0.007 &  0.483$\pm$0.006 &  172.1$\pm$0.3 &  11.8$\pm$0.2  &  0.54$_{0.29}^{0.59}$ &     0.34$_{0.16}^{0.59}$ & 0.00  & 0.51 & 0.17 & 0.32  & None              \\
UGC03253 &    0.0145  &      10.390  &      Sb    &  0.66    &     19.7   &      1.02    &    0.42   &      0.60$\pm$0.01   &  0.295$\pm$0.007 &  78.4$\pm$0.7  &  35.8$\pm$0.4  &  0.21$_{0.09}^{0.24}$ &     0.21$_{0.09}^{0.34}$ & 0.00  & 0.80 & 0.15 & 0.05  & Barlens            \\
UGC03995 &    0.0164  &      10.773  &      Sb    &  0.67    &     20.0   &      2.03    &    0.67   &      0.526$\pm$0.007 &  0.313$\pm$0.006 &  93.6$\pm$0.3  &  119.6$\pm$0.2 &  0.31$_{0.14}^{0.36}$ &     0.19$_{0.06}^{0.31}$ & 0.00  & 0.78 & 0.05 & 0.17  & Barlens            \\
NGC2486 &    0.0161   &      10.483  &      Sab   &  0.24    &     20.7   &      2.06    &    0.55   &      0.591$\pm$0.009 &  0.425$\pm$0.007 &  90.7$\pm$0.5  &  49.8$\pm$0.4  &  0.31$_{0.14}^{0.36}$ &     0.34$_{0.09}^{0.51}$ & 0.00  & 0.72 & 0.21 & 0.06  & Barlens            \\
UGC04145 &    0.0161  &      10.548  &      Sa    &  0.70    &     20.3   &      1.09    &    0.42   &      0.50$\pm$0.01   &  0.418$\pm$0.007 &  138.2$\pm$0.7 &  130.4$\pm$0.4 &  0.71$_{0.39}^{0.76}$ &     0.21$_{0.06}^{0.41}$ & 0.00  & 0.25 & 0.02 & 0.73  & None               \\
UGC04195 &    0.0170  &      10.186  &      Sb    &  0.51    &     20.4   &      1.23    &    0.53   &      0.590$\pm$0.009 &  0.259$\pm$0.007 &  13.2$\pm$0.5  &  151.0$\pm$0.4 &  0.19$_{0.09}^{0.21}$ &     0.19$_{0.06}^{0.29}$ & 0.00  & 0.83 & 0.13 & 0.04  & Barlens            \\
NGC2530  &    0.0174  &      9.988   &      Sd    &  0.30    &     20.6   &      1.86    &    1.03   &      0.84$\pm$0.01   &  0.33$\pm$0.01   &  156.3$\pm$0.7 &  38.9$\pm$0.4  &  0.26$_{0.14}^{0.29}$ &     0.59$_{0.19}^{1.01}$ & 0.10  & 0.24 & 0.65 & 0.01  & None               \\
NGC2543 &   0.0091    &      10.036  &      Sbc   &  0.61    &     20.1   &      1.39    &    0.38   &      0.583$\pm$0.007 &  0.348$\pm$0.006 &  46.0$\pm$0.3  &  91.8$\pm$0.2  &  0.24$_{0.09}^{0.26}$ &     0.26$_{0.09}^{0.41}$ & 0.00  & 0.77 & 0.18 & 0.04  & Barlens?           \\
UGC04308 &    0.0127  &      10.081  &      Sc    &  0.31    &     20.5   &      3.40    &    0.57   &      0.740$\pm$0.009 &  0.190$\pm$0.007 &  126.6$\pm$0.5 &  96.8$\pm$0.4  &  0.16$_{0.09}^{0.21}$ &     0.19$_{0.06}^{0.26}$ & 0.00  & 0.84 & 0.12 & 0.03  & None               \\
NGC2553 &    0.0165   &      10.665  &      Sb    &  0.79    &     20.2   &      1.66    &    0.72   &      0.59$\pm$0.01   &  0.511$\pm$0.007 &  59.5$\pm$0.7  &  11.3$\pm$0.4  &  0.36$_{0.16}^{0.39}$ &     0.41$_{0.14}^{0.66}$ & 0.00  & 0.61 & 0.34 & 0.04  & Barlens            \\
NGC2558 &    0.0174   &      10.722  &      Sb    &  0.66    &     20.1   &      1.61    &    0.65   &      0.67$\pm$0.01   &  0.429$\pm$0.007 &  157.4$\pm$0.7 &  167.8$\pm$0.4 &  0.46$_{0.26}^{0.61}$ &     0.39$_{0.11}^{0.51}$ & 0.00  & 0.49 & 0.10 & 0.41  & None               \\
NGC2565 &    0.0127   &      10.605  &      Sb    &  0.63    &     19.5   &      1.49    &    0.46   &      0.501$\pm$0.007 &  0.386$\pm$0.006 &  166.3$\pm$0.3 &  176.8$\pm$0.2 &  0.61$_{0.36}^{0.66}$ &     0.21$_{0.06}^{0.36}$ & 0.00  & 0.35 & 0.03 & 0.62  & BP?                \\
NGC2880 &   0.0066    &      10.308  &      E7    &  0.72    &     19.9   &      2.86    &    1.06   &      0.571$\pm$0.003 &  0.541$\pm$0.003 &  143.2$\pm$0.1 &  81.5$\pm$0.1  &  0.36$_{0.19}^{0.36}$ &     0.51$_{0.16}^{0.89}$ & 0.00  & 0.43 & 0.55 & 0.02  & None               \\
NGC3300 &    0.0117   &      10.475  &      S0a   &  0.77    &     19.1   &      1.25    &    0.40   &      0.551$\pm$0.007 &  0.420$\pm$0.006 &  174.6$\pm$0.3 &  43.7$\pm$0.2  &  0.26$_{0.11}^{0.29}$ &     0.34$_{0.09}^{0.51}$ & 0.00  & 0.71 & 0.25 & 0.04  & Barlens            \\
NGC3811 &    0.0119   &      10.321  &      Sbc   &  0.57    &     19.4   &      0.73    &    0.29   &      0.668$\pm$0.007 &  0.570$\pm$0.006 &  12.9$\pm$0.3  &  30.3$\pm$0.2  &  0.66$_{0.46}^{0.71}$ &     0.36$_{0.11}^{0.61}$ & 0.00  & 0.42 & 0.14 & 0.45  & None               \\
NGC4003 &    0.0235   &      10.946  &      S0a   &  0.75    &     20.8   &      1.44    &    1.51   &      0.70$\pm$0.01   &  0.271$\pm$0.007 &  175.4$\pm$0.7 &  143.6$\pm$0.4 &  0.24$_{0.11}^{0.29}$ &     0.24$_{0.09}^{0.36}$ & 0.00  & 0.78 & 0.15 & 0.08  & Barlens            \\
NGC4185 &    0.0148   &      10.427  &      Sbc   &  -0.43   &     20.2   &      1.01    &    0.66   &      0.666$\pm$0.004 &  0.796$\pm$0.003 &  166.9$\pm$0.3 &  172.1$\pm$0.2 &  0.94$_{0.84}^{0.96}$ &     0.64$_{0.49}^{0.76}$ & 0.00  & 0.09 & 0.04 & 0.88  & None               \\
NGC4210 &    0.0105   &      10.153  &      Sb    &  0.49    &     19.9   &      0.82    &    0.32   &      0.731$\pm$0.004 &  0.274$\pm$0.003 &  94.09$\pm$0.3 &  47.4$\pm$0.23 &  0.21$_{0.11}^{0.24}$ &     0.26$_{0.11}^{0.44}$ & 0.00  & 0.68 & 0.29 & 0.03  & Barlens            \\
NGC5157 &    0.0263   &      11.021  &      Sab   &  0.68    &     20.2   &      1.21    &    1.61   &      0.78$\pm$0.01   &  0.320$\pm$0.007 &  105.7$\pm$0.7 &  144.1$\pm$0.4 &  0.26$_{0.11}^{0.31}$ &     0.34$_{0.11}^{0.54}$ & 0.00  & 0.61 & 0.35 & 0.04  & None                \\
NGC5205 &   0.0075    &      9.8618  &      Sbc   &  0.53    &     19.6   &      0.86    &    0.28   &      0.63$\pm$0.01   &  0.335$\pm$0.007 &  152.6$\pm$0.7 &  104.7$\pm$0.4 &  0.24$_{0.11}^{0.26}$ &     0.26$_{0.09}^{0.44}$ & 0.00  & 0.72 & 0.24 & 0.03  & BP                  \\
UGC08781 &    0.0274  &      10.871  &      Sb    &  0.63    &     20.2   &      2.13    &    0.99   &      0.60$\pm$0.01   &  0.274$\pm$0.007 &  165.9$\pm$0.7 &  172.6$\pm$0.4 &  0.36$_{0.14}^{0.46}$ &     0.21$_{0.09}^{0.31}$ & 0.00  & 0.62 & 0.04 & 0.34  & None                \\
NGC5378 &    0.0121   &      10.140  &      Sb    &  0.17    &     20.9   &      2.51    &    0.82   &      0.74$\pm$0.03   &  0.31$\pm$0.03   &  80$\pm$2.3    &  40.0$\pm$1.7  &  0.24$_{0.11}^{0.31}$ &     0.34$_{0.11}^{0.49}$ & 0.00  & 0.67 & 0.29 & 0.04  & Barlens              \\
NGC5406  &   0.0192   &      11.017  &      Sb    &  0.40    &     19.9   &      1.30    &    0.63   &      0.771$\pm$0.004 &  0.293$\pm$0.003 &  120.0$\pm$0.3 &  53.5$\pm$0.3  &  0.19$_{0.11}^{0.24}$ &     0.64$_{0.21}^{0.84}$ & 0.00  & 0.27 & 0.73 & 0.01  & None                \\
NGC5520 &   0.0081    &      9.7922  &      Sbc   &  0.44    &     19.5   &      ---     &    ---    &      0.56$\pm$0.01   &  0.531$\pm$0.006 &  65$\pm$1.0    &  65.3$\pm$0.5  &  1.00$_{0.94}^{1.00}$ &     0.24$_{0.04}^{0.36}$ & 0.00  & 0.02 & 0.00 & 0.98  & None                  \\
NGC5519  &  0.0271    &      10.660  &      Sb    &  0.49    &     20.3   &      3.24    &    2.12   &      0.71$\pm$0.02   &  0.46$\pm$0.02   &  75.9$\pm$0.9  &  105.9$\pm$0.8 &  0.44$_{0.21}^{0.49}$ &     0.36$_{0.11}^{0.61}$ & 0.00  & 0.58 & 0.26 & 0.16  & None                  \\
IC0994 &    0.0269    &      11.072  &      S0a   &  0.80    &     19.8   &      2.16    &    0.87   &      0.51$\pm$0.01   &  0.481$\pm$0.007 &  14.6$\pm$0.7  &  29.2$\pm$0.4  &  0.59$_{0.31}^{0.66}$ &     0.34$_{0.09}^{0.49}$ & 0.00  & 0.41 & 0.05 & 0.54  & None                   \\
NGC5602 &   0.0092    &      10.322  &      S0a   &  4.92    &     20.8   &      2.39    &    0.46   &      0.52$\pm$0.03   &  0.59$\pm$0.03   &  167$\pm$2.3   &  165.1$\pm$1.7 &  0.94$_{0.81}^{1.00}$ &     0.41$_{0.26}^{0.51}$ & 0.00  & 0.05 & 0.01 & 0.94  & None                   \\
UGC09291 &    0.0116  &      9.6661  &      Scd   &  0.44    &     20.4   &      1.06    &    0.41   &      0.514$\pm$0.009 &  0.319$\pm$0.007 &  107.2$\pm$0.5 &  73.0$\pm$0.4  &  0.24$_{0.09}^{0.29}$ &     0.21$_{0.06}^{0.31}$ & 0.00  & 0.82 & 0.08 & 0.10  & None                  \\
NGC5735  &    0.0145  &       10.127 &      Sbc   &  0.45    &     20.3   &      1.20    &    0.42   &      0.90$\pm$0.01   &  0.223$\pm$0.007 &  32.6$\pm$0.7  &  94.9$\pm$0.4  &  0.16$_{0.06}^{0.21}$ &     0.64$_{0.21}^{0.96}$ & 0.02  & 0.19 & 0.79 & 0.00  & None                 \\
UGC09492 &    0.0299  &      11.082  &      Sab   &  0.78    &     20.1   &      1.58    &    1.37   &      0.45$\pm$0.01   &  0.420$\pm$0.007 &  54.2$\pm$0.7  &  80.8$\pm$0.4  &  0.36$_{0.16}^{0.44}$ &     0.26$_{0.06}^{0.39}$ & 0.00  & 0.77 & 0.06 & 0.17  & Barlens?             \\
IC4534 &    0.0185    &      10.723  &      S0a   &  0.48    &     20.0   &      2.10    &    0.85   &      0.564$\pm$0.009 &  0.293$\pm$0.007 &  162.9$\pm$0.5 &  179.1$\pm$0.4 &  0.31$_{0.16}^{0.44}$ &     0.24$_{0.06}^{0.31}$ & 0.00  & 0.68 & 0.04 & 0.28  & BP                   \\
NGC5876 &    0.0126   &      10.678  &      S0a   &  0.69    &     20.6   &      1.57    &    0.72   &      0.433$\pm$0.004 &  0.483$\pm$0.003 &  52.4$\pm$0.3  &  179.4$\pm$0.2 &  0.24$_{0.11}^{0.26}$ &     0.31$_{0.11}^{0.48}$ & 0.00  & 0.76 & 0.22 & 0.02  & Barlens              \\
NGC5888 &    0.0308   &      11.135  &      Sb    &  0.61    &     19.8   &      1.40    &    1.06   &      0.596$\pm$0.009 &  0.243$\pm$0.007 &  153.2$\pm$0.5 &  170.5$\pm$0.4 &  0.31$_{0.14}^{0.34}$ &     0.14$_{0.09}^{0.26}$ & 0.00  & 0.75 & 0.05 & 0.20  & BP                   \\
NGC6154 &    0.0216   &      10.744  &      Sab   &  0.64    &     20.5   &      2.32    &    1.15   &      0.79$\pm$0.01   &  0.250$\pm$0.007 &  135.9$\pm$0.7 &  135.4$\pm$0.4 &  0.24$_{0.11}^{0.31}$ &     0.26$_{0.11}^{0.36}$ & 0.00  & 0.79 & 0.14 & 0.08  & Barlens?             \\
NGC6278 &    0.0111   &      10.719  &      S0a   &  0.69    &     20.0   &      2.31    &    0.64   &      0.530$\pm$0.004 &  0.379$\pm$0.003 &  126.8$\pm$0.3 &  116.0$\pm$0.2 &  0.54$_{0.34}^{0.64}$ &     0.26$_{0.09}^{0.36}$ & 0.00  & 0.42 & 0.03 & 0.55  & None                 \\
NGC6497 &    0.0217   &      10.865  &      Sab   &  0.58    &     20.1   &      3.33    &    1.11   &      0.495$\pm$0.009 &  0.392$\pm$0.007 &  111.2$\pm$0.5 &  160.8$\pm$0.4 &  0.21$_{0.09}^{0.26}$ &     0.29$_{0.09}^{0.41}$ & 0.00  & 0.79 & 0.18 & 0.03  & Barlens              \\
UGC11228 &    0.0206  &      10.867  &      S0    &  0.81    &     20.3   &      2.50    &    1.14   &      0.67$\pm$0.01   &  0.286$\pm$0.007 &  176.2$\pm$0.7 &  157.5$\pm$0.4 &  0.31$_{0.14}^{0.36}$ &     0.21$_{0.11}^{0.34}$ & 0.00  & 0.77 & 0.08 & 0.14  & None                 \\
NGC6941 &    0.0216   &      10.885  &      Sb    &  0.66    &     20.2   &      1.76    &    0.85   &      0.72$\pm$0.01   &  0.248$\pm$0.007 &  130.5$\pm$0.7 &  109.1$\pm$0.4 &  0.26$_{0.14}^{0.29}$ &     0.19$_{0.11}^{0.31}$ & 0.00  & 0.79 & 0.12 & 0.09  & BP?                  \\
NGC6945 &    0.0136   &      10.946  &      S0    &  0.74    &     19.9   &      3.74    &    1.09   &      0.649$\pm$0.007 &  0.507$\pm$0.006 &  120.2$\pm$0.3 &  86.7$\pm$0.2  &  0.46$_{0.21}^{0.51}$ &     0.36$_{0.14}^{0.64}$ & 0.00  & 0.60 & 0.24 & 0.16  & Barlens?             \\
NGC7321 &    0.0238   &      10.691  &      Sbc   &  0.59    &     19.3   &      1.79    &    0.53   &      0.65$\pm$0.01   &  0.412$\pm$0.007 &  22.2$\pm$0.7  &  71.7$\pm$0.4  &  0.29$_{0.14}^{0.34}$ &     0.39$_{0.11}^{0.61}$ & 0.00  & 0.60 & 0.36 & 0.04  & Barlens              \\
NGC7563 &    0.0139   &      10.720  &      Sa    &  0.71    &     21.5   &      2.47    &    1.21   &      0.513$\pm$0.009 &  0.392$\pm$0.007 &  151.5$\pm$0.5 &  91.9$\pm$0.4  &  0.21$_{0.09}^{0.24}$ &     0.29$_{0.11}^{0.51}$ & 0.00  & 0.66 & 0.33 & 0.01  & Barlens             \\
NGC7591 &    0.0164   &      10.610  &      Sbc   &  0.68    &     20.3   &      2.30    &    2.00   &      0.48$\pm$0.01   &  0.484$\pm$0.007 &  148.4$\pm$0.7 &  10.5$\pm$0.4  &  0.31$_{0.14}^{0.34}$ &     0.31$_{0.09}^{0.48}$ & 0.00  & 0.78 & 0.17 & 0.05  & None                \\
NGC7611 &    0.0109   &      10.721  &      S0    &  0.84    &     18.8   &      1.66    &    0.28   &      0.487$\pm$0.007 &  0.711$\pm$0.006 &  137.0$\pm$0.3 &  1.7$\pm$0.2   &  0.51$_{0.26}^{0.54}$ &     0.56$_{0.19}^{0.81}$ & 0.00  & 0.63 & 0.33 & 0.04  & None                \\
NGC7623 &    0.0125   &      10.565  &      S0    &  0.90    &     20.6   &      1.90    &    1.02   &      0.719$\pm$0.009 &  0.400$\pm$0.007 &  7.0$\pm$0.5   &  161.1$\pm$0.4 &  0.39$_{0.19}^{0.46}$ &     0.34$_{0.09}^{0.54}$ & 0.00  & 0.64 & 0.21 & 0.15  & None                \\
NGC7671 &    0.0137   &      10.655  &      S0    &  0.88    &     18.8   &      1.76    &    0.48   &      0.596$\pm$0.004 &  0.617$\pm$0.003 &  135.1$\pm$0.3 &  24.0$\pm$0.2  &  0.46$_{0.29}^{0.49}$ &     0.66$_{0.26}^{1.21}$ & 0.31  & 0.26 & 0.41 & 0.01  & None                \\ 
NGC7691 &    0.0134   &      10.176  &      Sbc   &  0.52    &     20.5   &      1.38    &    0.56   &      0.78$\pm$0.01   &  0.361$\pm$0.007 &  160.0$\pm$0.7 &  2.7$\pm$0.4   &  0.36$_{0.16}^{0.42}$ &     0.31$_{0.11}^{0.54}$ & 0.00  & 0.62 & 0.27 & 0.11  & None                \\
NGC7716 &    0.0087   &      10.289  &      Sb    &  0.52    &     19.7   &      2.87    &    0.31   &      0.822$\pm$0.008 &  0.514$\pm$0.002 &  39.4$\pm$0.4  &  15.7$\pm$0.2  &  0.49$_{0.29}^{0.56}$ &     0.51$_{0.16}^{0.81}$ & 0.00  & 0.45 & 0.39 & 0.16  & None                \\
UGC04455 &    0.0310  &      10.878  &      Sb    & -0.20    &     20.3   &      0.97    &    0.81   &      0.74$\pm$0.03   &  0.23$\pm$0.03   &  11$\pm$2.3    &  165$\pm$1.7   &  0.21$_{0.09}^{0.26}$ &     0.21$_{0.06}^{0.33}$ & 0.00  & 0.80 & 0.14 & 0.06  & Barlens?            \\
UGC05187 &    0.0049  &      8.7767  &      Sdm   &  0.05    &     20.6   &      ---     &    ---    &      0.825$\pm$0.008 &  0.392$\pm$0.008 &  133.8$\pm$1.6 &  79.9$\pm$0.7  &  0.31$_{0.14}^{0.36}$ &     0.59$_{0.19}^{0.94}$ & 0.01  & 0.32 & 0.65 & 0.02  & None                 \\
UGC05377 &    0.0071  &      8.8268  &      Sdm   &  0.39    &     21.1   &      ---     &    ---    &      0.835$\pm$0.008 &  0.321$\pm$0.006 &  77.6$\pm$1.5  &  135.1$\pm$0.7 &  0.24$_{0.11}^{0.29}$ &     0.59$_{0.16}^{0.86}$ & 0.00  & 0.32 & 0.67 & 0.01  & None                 \\
UGC07129 &   0.0040   &      9.2663  &      Sab   & -0.06    &     19.6   &      ---     &    ---    &      0.64$\pm$0.01   &  0.164$\pm$0.006 &  73$\pm$1.0    &  61.6$\pm$0.5  &  0.19$_{0.09}^{0.24}$ &     0.14$_{0.06}^{0.19}$ & 0.00  & 0.82 & 0.04 & 0.14  & None                 \\
NGC0495 &    0.0135   &      10.902  &      Sab   & -0.81    &     19.8   &      1.58    &    0.67   &      0.73$\pm$0.01   &  0.345$\pm$0.007 &  10.4$\pm$0.7  &  10.5$\pm$0.4  &  0.41$_{0.16}^{0.46}$ &     0.26$_{0.14}^{0.46}$ & 0.00  & 0.63 & 0.13 & 0.23  & None                 \\
KUG1349+143 &  0.0247 &      10.076  &      Sbc   & -0.05    &     20.9   &      1.65    &    0.26   &      0.63$\pm$0.03   &  0.23$\pm$0.04   &  96$\pm$2.3    &  96$\pm$1.3    &  0.29$_{0.11}^{0.41}$ &     0.19$_{0.06}^{0.29}$ & 0.00  & 0.71 & 0.04 & 0.24  & None                 \\
IC1078 &    0.0287    &      10.645  &      Sb    &  0.08    &     20.7   &      1.92    &    0.60   &      0.82$\pm$0.02   &  0.27$\pm$0.02   &  99.5$\pm$0.9  &  100.1$\pm$0.8 &  0.26$_{0.11}^{0.34}$ &     0.29$_{0.11}^{0.44}$ & 0.00  & 0.72 & 0.21 & 0.07  & None                 \\
NGC0515 &   0.0170    &      11.038  &      E7    & -0.25    &     19.9   &      1.77    &    0.54   &      0.77$\pm$0.01   &  0.466$\pm$0.007 &  103.9$\pm$0.7 &  131.9$\pm$0.4 &  0.46$_{0.26}^{0.51}$ &     0.36$_{0.11}^{0.66}$ & 0.00  & 0.54 & 0.32 & 0.15  & Barlens?             \\
NGC2780 &   0.0068    &      9.5927  &      Sbc   &  -0.17   &     20.2   &      0.56    &    0.16   &      0.721$\pm$0.009 &  0.166$\pm$0.007 &  143.7$\pm$0.5 &  113.9$\pm$0.4 &  0.14$_{0.06}^{0.19}$ &     0.16$_{0.06}^{0.21}$ & 0.00  & 0.85 & 0.10 & 0.05  & None                 \\
UGC06517 &  0.0084    &      9.3464  &      Sc    &  0.16    &     19.5   &      ---     &    ---    &      0.60$\pm$0.01   &  0.287$\pm$0.008 &  34.4$\pm$1.5  &  105.6$\pm$0.8 &  0.16$_{0.06}^{0.16}$ &     0.36$_{0.11}^{0.64}$ & 0.00  & 0.39 & 0.61 & 0.00  & None                 \\
UGC12250 &    0.0240  &      11.106  &      Sbc   &  0.22    &     20.7   &      1.15    &    0.71   &      0.626$\pm$0.009 &  0.320$\pm$0.007 &  12.7$\pm$0.5  &  13.3$\pm$0.4  &  0.34$_{0.19}^{0.54}$ &     0.31$_{0.09}^{0.36}$ & 0.00  & 0.55 & 0.05 & 0.40  & None                 \\
NGC5947 &    0.0198   &      10.559  &      Sbc   &  0.39    &     20.6   &      2.71    &    0.61   &      0.811$\pm$0.009 &  0.294$\pm$0.007 &  63.7$\pm$0.5  &  25.7$\pm$0.4  &  0.24$_{0.11}^{0.29}$ &     0.36$_{0.14}^{0.54}$ & 0.00  & 0.59 & 0.38 & 0.04  & Barlens              \\
UGC03552 &    0.0160  &      10.067  &      Sd    &  0.46    &     20.3   &      ---     &    ---    &      0.46$\pm$0.01   &  0.475$\pm$0.008 &  78.3$\pm$1.5  &  96.3$\pm$0.8  &  0.51$_{0.24}^{0.59}$ &     0.34$_{0.11}^{0.46}$ & 0.00  & 0.57 & 0.05 & 0.38  & None                 \\
NGC3323 &    0.0173   &      9.8078  &      Scd   &  0.39    &     19.3   &      ---     &    ---    &      0.495$\pm$0.008 &  0.161$\pm$0.006 &  157.1$\pm$1.5 &  163.9$\pm$0.7 &  0.21$_{0.09}^{0.34}$ &     0.14$_{0.06}^{0.16}$ & 0.00  & 0.63 & 0.01 & 0.36  & None                 \\
NGC2767 &    0.0165   &      10.750  &      S0    &  0.83    &     18.9   &      1.22    &    0.33   &      0.733$\pm$0.009 &  0.417$\pm$0.007 &  169.6$\pm$0.5 &  150.7$\pm$0.4 &  0.44$_{0.24}^{0.51}$ &     0.34$_{0.11}^{0.54}$ & 0.00  & 0.59 & 0.19 & 0.22  & None                 \\
\hline                                 
%\end{tabular}
\begin{minipage}{215mm}
Note.  (1)  Galaxy name;  (2), (3), and  (4) redshift,  galaxy stellar
mass,  and  galaxy  Hubble   type  from  \citet{walcher14};  (5),  (6)
integrated  disc $g-r$  color and  central surface  brightness of  the
disc; (7) and (8) S\'ersic index  and effective radius (in kpc) of the
bulge; (9)  and (10) apparent  axial ratios of  the disc and  the bar,
(11) and (12)  position angle of the  disc and the bar,  (13) and (14)
most probable intrinsic  axial ratios $B/A$ and $C/A$ of  the bar, the
1$\sigma$ probability values for each case are also shown; (15), (16),
(17)  and  (18) probability  for  a  given  bar to  prolate  off-plane
(pro. o), prolate  in-plane (pro. i) , oblate off-plane  (obl. o), and
oblate in-plane (obl. i),  respectively; (19) visual classification to
detect inner boxy/peanuts  structures (see text for  an explanation of
the different classes). Columns  (5, 6, 7, 8, 9, 10,  11, 12) are from
\citep{mendezabreu17}.
\end{minipage}
\end{longtable}
\end{center}
\end{tiny}
\end{landscape}

%-----------------------------------------------------
\twocolumn
\section{Influence of boxy/peanut structures in the photometric decomposition of bars}
\label{sec:barlens}

We devised a  set of galaxy image simulations to  derive the influence
of a  boxy/peanut inner component  in the geometric parameters  of the
bars obtained from our photometric decomposition. To this aim, we used
a   similar   approach   to   that  described   in   Sect.    5.2   of
\citet{mendezabreu17}. We  generated 500  mock galaxies composed  by a
bulge (described using a S\'ersic distribution), a disc (exponential),
an outer  bar (Ferrers), and  an inner boxy/peanut  structure modelled
with  another Ferrers  profile.  We  considered the  inner boxy/peanut
structure  to be  the same  structure  as the  barlenses described  in
\citet{laurikainen11,laurikainen13}.  The  use of either a  Ferrers or
S\'ersic   profile  to   describe  the   barlens  surface   brigthness
distribution was mentioned by \citet{athanassoula15} as giving similar
results.  We  opted for a Ferrers  profile since it was  more directly
comparable with the available observational constraints for barlenses,
in particular to  create a model with given ratios  between the length
of the outer and inner regions of the bar. The range of values used to
build  the   bulge,  disc,   and  bar   components  were   taken  from
\citet{mendezabreu17},  including  $q_{\rm  disk}$= [0.55,1]  and $q_{\rm
  bar}=[0.2,0.4]$.    They  are   representative  of   the  structural
parameters in the  $i-$band for barred galaxies present  in the CALIFA
survey.   The values  of  the  Ferrers profile  used  to describe  the
barlens  structure  were  extracted  from  \citet{athanassoula15}  and
\citet{laurikainensalo17}.    We  used   random  values   between  the
following  ranges:  $L_{\rm  bl}/L_{\rm bar}  =  [0.31,1.3]$,  $r_{\rm
  bl}/r_{\rm  bar}  = [0.4,0.8]$,  $q_{\rm  bl}  = [0.5,1]$,  PA$_{\rm
  bl}$=PA$_{\rm bar}$.

\begin{figure*}
\includegraphics[width=0.9\textwidth]{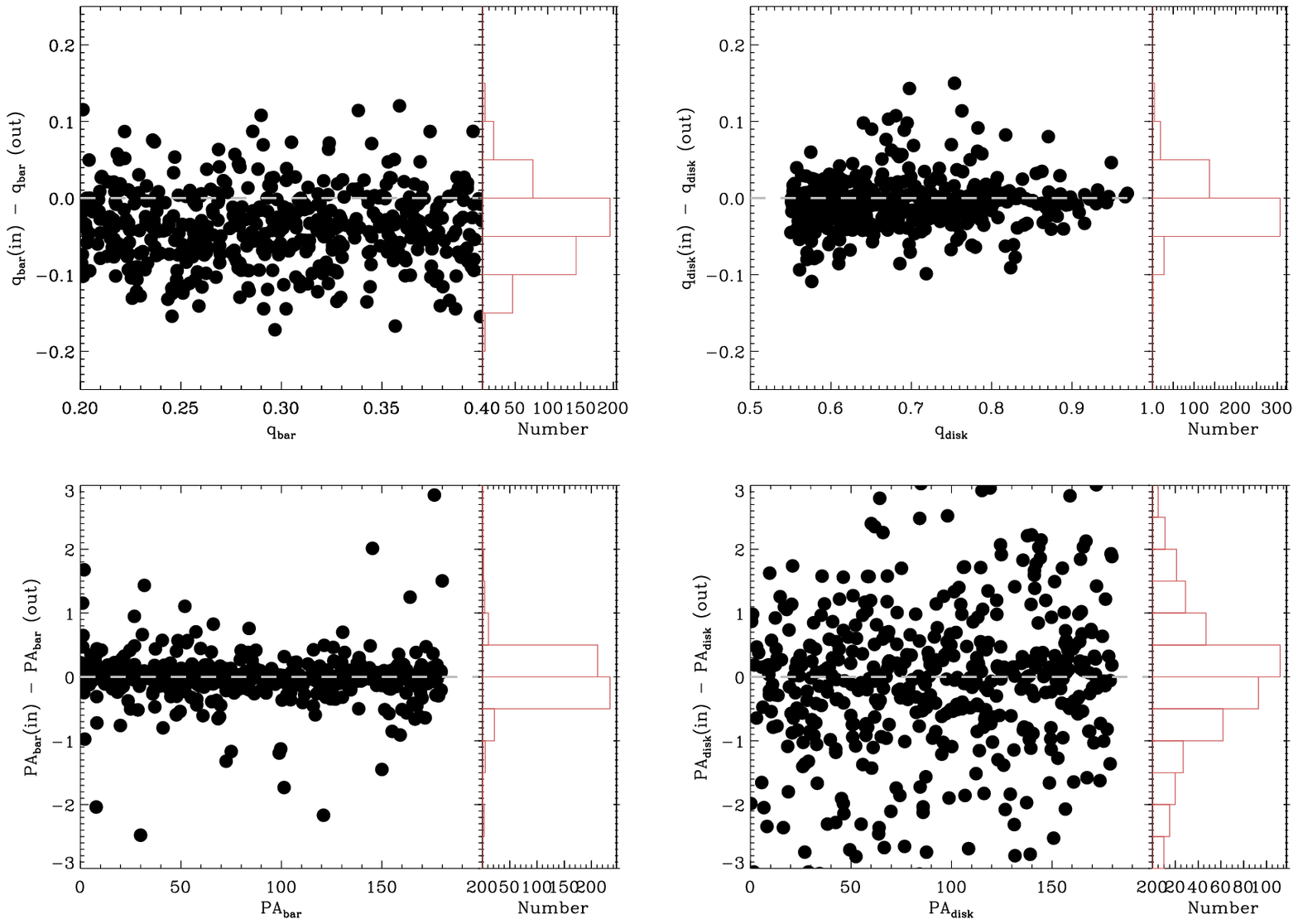}
\caption{Differences between the input and output values obtained from
  our mock  galaxy simulations including barlenses  for the parameters
  involved in  our analysis:  bar axis ratio  (upper left),  disc axis
  ratio (upper right), bar position angle (bottom left), disc position
  angle (bottom  right). The  histograms of  the differences  are also
  shown.}
\label{fig:ellipsimu}
\end{figure*}

The mock galaxies  were analysed as if they were  pure barred galaxies
without any inner  boxy/peanut component, i.e., only a  bulge, a disc,
and the  outer parts of  the bar were  considered to model  the galaxy
light.   This allowed  us to  quantify the  effect of  the boxy/peanut
surface brightness  distribution on  the photometric  decomposition of
our barred  galaxies, in particular,  the effect of  the ellipticities
and position  angles.  Fig.~\ref{fig:ellipsimu} shows  the differences
in the  four structural parameters  involved in our  analysis ($q_{\rm
  bar}$, $q_{\rm disc}$,  PA$_{\rm bar}$, PA$_{\rm disc}$)  due to the
addition of a barlens structure in  the galaxy centre.  We compute the
mean (systematic) and  rms (statistical) errors between  the input and
output values  of the  mock galaxies  obtaining that:  $\langle q_{\rm
  bar}  $(in) -  $q_{\rm bar}$  (out)$\rangle$ =  -0.037 $\pm$  0.048,
$\langle q_{\rm disc}  $(in) - $q_{\rm disc}  $(out)$\rangle$ = -0.007
$\pm$  0.031  ,  $\langle  $PA$_{\rm   bar}$  (in)  -  PA$_{\rm  bar}$
(out)$\rangle$ =  -0.03 $\pm$ 0.51,  $\langle $PA$_{\rm disc}$  (in) -
PA$_{\rm  disc}$  (out)$\rangle$  =   0.14  $\pm$  2.25.  Despite    the fact
  that the
statistical errors are always larger  than any possible systematic, we
found a  weak bias of  the bar axis  ratio towards larger  values when
including the barlens component.  This can  be expected if part of the
barlens surface  brightness, which  is always  rounder than  the outer
bar, is incoporated into the bar component.

%--------------------------------------------------------
%--------------------------------------------------------

\begin{figure*}
\includegraphics[width=0.9\textwidth]{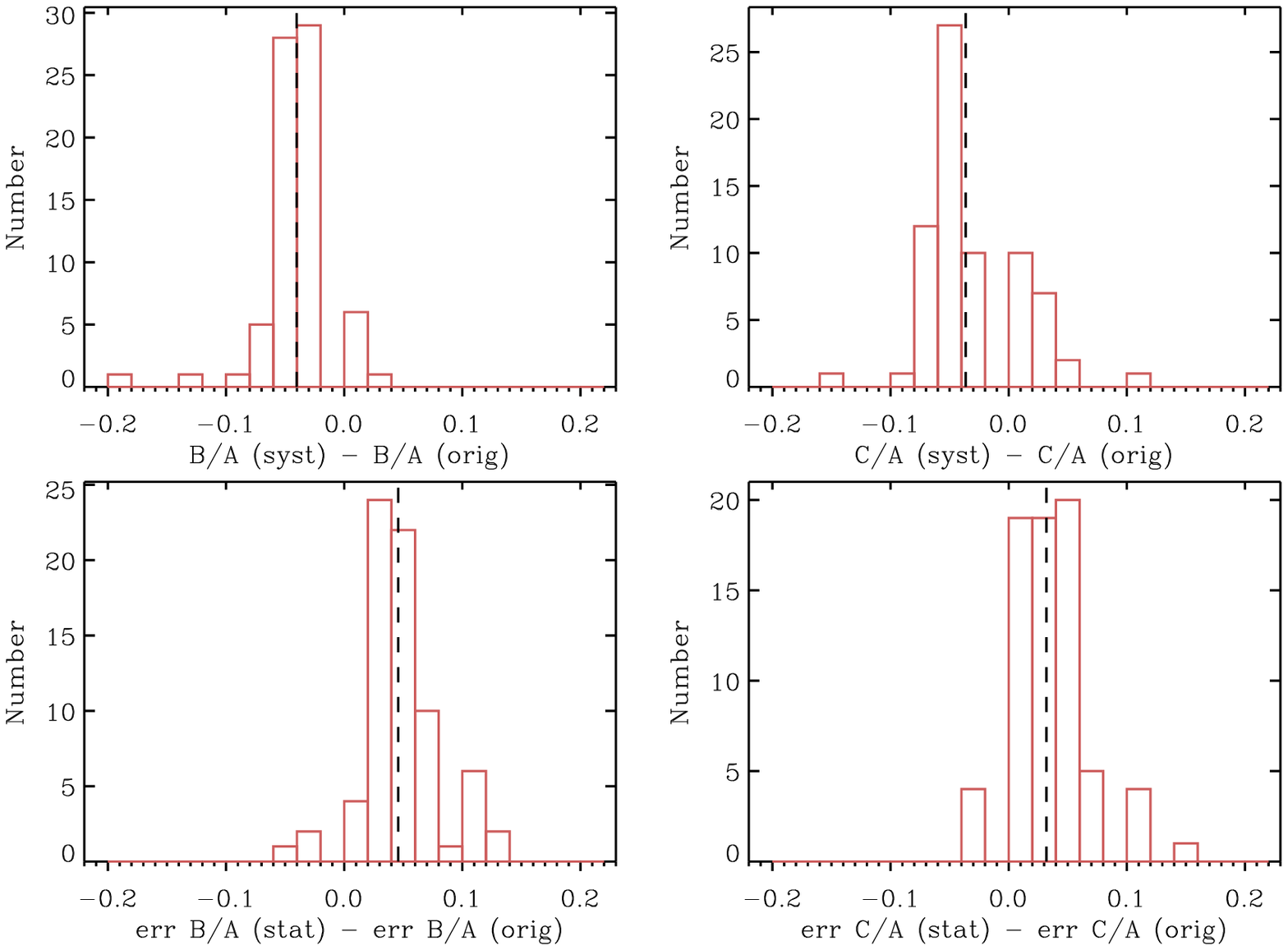}
\vspace{0.3cm}
\caption{{\it Top  panels.}  Distribution  of differences  between the
  derived values of  the intrinsic $B/A$ and $C/A$  of bars correcting
  for the systematic (syst) error  on the observational parameters due
  to the  presence of a barlens  structure and those obtained  in this
  paper  (orig). {\it  Bottom  panels.}   Distribution of  differences
  between the derived  uncertainties in the intrinsic  $B/A$ and $C/A$
  semiaexes  of bars  assuming  the statistical  (stat)  error on  the
  observational parameters due to the  presence of a barlens structure
  and those obtained in this paper (orig).}
\label{fig:comp_barlens}
\end{figure*}

At this point,  we study how the previous errors  in the observational
measurements propagate into the derived  intrinsic shape of bulges. We
used both  the systematic and  statistical errors to  understand their
effect into the  intrinsic $B/A$ and $C/A$ semiaxis  ratios. First, we
recomputed the bar intrinsic shape (using the methodology described in
Sect.~\ref{sec:bars3D})   correcting    for   the    mean   deviations
(systematic) in the observational parameters obtained previosuly. This
will provide  us with the  typical difference  in the $B/A$  and $C/A$
values assuming that all galaxies host a barlens structure. Second, we
recomputed  the typical  uncertainty  in the  bar  intrinsic shape  by
assuming the photometric errors in  the observational parameters to be
the statistical  errors computed  using the barlens  simulations. This
help us to  quantify how the uncertainties in the  intrinsic $B/A$ and
$C/A$ semiaxes  are affected by  the presence of a  barlens structure.
Fig.~\ref{fig:comp_barlens}  show the  results  of  this analysis.  We
found that,  if a galaxy  have a barlens  structure that has  not been
included in  the photometric  decomposition procedure, both  $B/A$ and
$C/A$  would  be  systematically   overestimated  by  0.04  and  0.04,
respectively.   Similarly, the  uncertainties in  the intrinsic  shape
derived by our method  (Sect.~\ref{sec:bars3D}) would be understimated
by 0.05 and 0.03 in $B/A$ and $C/A$, respectively.  Therefore, even if
the results presented  in this paper are not affected  by the possible
presence of  an inner boxy/peanut  structure (characterised here  as a
barlens), small variations can be found.

%--------------------------------------------------------
%--------------------------------------------------------

% Don't change these lines
\bsp	% typesetting comment
\label{lastpage}
\end{document}